\documentclass[12pt]{article}
\RequirePackage[authoryear]{natbib}

\usepackage[margin=1in]{geometry}
\usepackage{amsmath}
\usepackage{natbib}
\usepackage{amssymb}
\usepackage{bbm}
\usepackage{tikz}
\usetikzlibrary{shapes,arrows}
\usetikzlibrary{decorations.pathreplacing}
\usetikzlibrary{positioning}
\usepackage{graphicx}
\usepackage{multirow}
\usepackage{enumitem}
\usepackage{booktabs,caption,subcaption}
\usepackage{url}

\newtheorem{theorem}{Theorem}

\newcommand{\bg}{\mathbf{g}}
\newcommand{\bG}{\mathbf{G}}

\newcommand{\bS}{\mathbf{S}}
\newcommand{\bY}{\mathbf{Y}}

\newcommand{\bX}{\mathbf{X}}
\newcommand{\bZ}{\mathbf{Z}}
\newcommand{\0}{\mathbf{0}}

\newcommand{\mA}{\mathbf{A}}
\newcommand{\mC}{\mathbf{C}}
\newcommand{\mD}{\mathbf{D}}
\newcommand{\mDT}{\mathbf{\Tilde{D}}}
\newcommand{\mH}{\mathbf{H}}
\newcommand{\mI}{\mathbf{I}}
\newcommand{\mK}{\mathbf{K}}
\newcommand{\mR}{\mathbf{R}}
\newcommand{\mRT}{\mathbf{\Tilde{R}}}
\newcommand{\mV}{\mathbf{V}}
\newcommand{\mW}{\mathbf{W}}
\newcommand{\mX}{\mathbf{X}} 
\newcommand{\mZ}{\mathbf{Z}}

\newcommand{\R}{\mathbb{R}}
\newcommand{\E}{\mathbb{E}}
\newcommand{\bbeta}{\boldsymbol\beta}
\newcommand{\bgamma}{\boldsymbol\gamma}

\newcommand{\boldeta}{\boldsymbol\eta}
\newcommand{\btheta}{\boldsymbol\theta}
\newcommand{\bTheta}{\boldsymbol\Theta}
\newcommand{\bxi}{\boldsymbol\xi}

\newcommand{\iSigma}{\mathbf{\Sigma}}
\newcommand{\bmu}{\boldsymbol\mu}

\allowdisplaybreaks

\begin{document}

\title{\bf Functional Regression with Intensively Measured Longitudinal Outcomes: A New Lens through Data Partitioning}
\author{Cole Manschot and Emily C. Hector\hspace{.2cm}\\
Department of Statistics, North Carolina State University}
\date{}
\maketitle
 
 \bigskip
\begin{abstract}
Estimation and inference with modern longitudinal data from wearable devices, which consist of biological signals at high-frequency time points, is burdened by massive computational costs. 
We propose a distributed estimation and inference procedure that efficiently estimates both functional and scalar parameters with intensively measured longitudinal outcomes. The procedure overcomes computational difficulties through a scalable divide-and-conquer algorithm that partitions the outcomes into smaller sets. We circumvent traditional basis selection problems by analyzing data using quadratic inference functions in smaller subsets such that the basis functions have a low dimension. To address the challenges of combining estimates from dependent subsets, we propose a statistically efficient one-step estimator derived from a constrained generalized method of moments objective function with a smoothing penalty. We show theoretically and numerically that the proposed estimator is as statistically efficient as non-distributed alternative approaches and more efficient computationally. We demonstrate the practicality of our approach with the analysis of accelerometer data from the National Health and Nutrition Examination Survey.
\end{abstract}

\noindent%
{\it Keywords:} Constrained optimization, Divide and conquer, Generalized method of moments, Parallel computing, Penalized regression.
\vfill

\section{Introduction}
\label{s:intro}

\begin{figure}[!hb]
    \centering
    \includegraphics[scale=1]{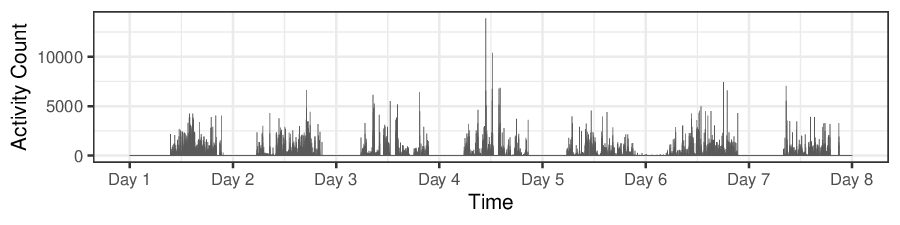}
    \caption{Total activity count by minute for a single participant over one week's time in the NHANES from 2003-2004 cohort.}
    \label{fig:nhanes_one_person}
\end{figure}

The recent explosion in the use of wearable devices to track well-being at an individual level offers unprecedented opportunities to monitor, assess and shape human health. 
A broad spectrum of publicly available devices including watches (e.g. Garmin), accessories (\={O}ura ring) and medical monitors (Freestyle Libre) 
measure physiological responses such as physical activity intensively over time, resulting in tens of thousands of outcomes per study participant. 
A canonical example, the National Health and Nutrition Examination Survey  \citep[NHANES,][]{nhanes_cdc} has measured physical activity for participants in the United States using hip-worn accelerometers since 2003. Activity counts for one participant are visualized in Figure \ref{fig:nhanes_one_person}. 
The Centers for Disease Control (CDC) continue to make available accelerometer data from the NHANES for each biannual cohort along with a wealth of individual phenotypic and health related data. 
Prior to the increased ease of access provided by the \verb|rnhanesdata| R package  \citep{leroux2019}, 
at least 170 studies had analyzed some measure of the accelerometer data from NHANES \citep{fuzeki2017}. 
A possible model, 
\begin{align} \label{eq:nhanesMean}
     &\log [ \E \{ \mathrm{Activity~Count}_i(t) | \mbox{Age}_i, \mbox{Sex}_i, \mbox{Mobility~Problem}_i, \mbox{BMI}_i \} ]  \notag \\
     &=  \eta_1 + \eta_2\mathrm{Age}_i + \eta_3\mathrm{Sex}_i + \eta_4 \mathrm{Mobility~Problem}_i + \beta_1(t) \mathrm{BMI}_i,
\end{align}
for participants $i=1, \ldots, N$, models expected activity count, a measure of minute-by-minute movement intensity, as a function of age, sex (1 indicating male), mobility problem (1 indicating presence of mobility problems) and body mass index (BMI).
Understanding individuals' physical activity and its association with BMI can offer rich insights into the association between activity levels and circadian rhythms \citep{youngstedt2019}, 
metabolism \citep{vieira2016}, cognitive abilities \citep{Wheeler2020}, and other correlated measures of health  \citep{stutz2019_sleep}. 

Existing methods for inference on $\eta_1, \eta_2, \eta_3, \eta_4$ and $\beta_1(t)$ are computationally burdened by the size of the data, with $10080$ outcomes observed on at least $2000$ individuals. Due to the data's size and complexity, models and analyses typically use crude summaries and averages, such as total activity count, ignoring the longitudinal nature of observations.
A few methods have been developed that use the full data in a regression-based approach
\citep{bai2018, schrack2014, xiao2015, goldsmith2015}.
The most widely used tool, functional regression, estimates a smooth functional relationship between intensively measured outcomes and covariates, and is burdened with tremendous computational costs due to the size of the data and complexity of the model. In this paper, we propose a theoretical framework for computationally and statistically efficient estimation and inference of these parameters based on a divide-and-conquer strategy.

As observed by \cite{liang1986}, use of likelihood-based methods can prove challenging with non-Gaussian models due to frequent reliance on copulas that are complicated to fit and interpret.  
The generalized estimating equations proposed by \cite{liang1986} require estimation of the covariance parameters and inversion of high-dimensional covariance matrices; in the case of the NHANES data, the covariance matrix has dimension $10080 \times 10080$ and its inversion is computationally prohibitive in iterative optimization \citep{song2007, banerjee2008}.
To avoid estimation of covariance parameters, quadratic inference functions \citep[QIF,][]{qulindsayli2000} model the inverse of the working correlation matrix as a linear expansion of known basis matrices, and show how the estimation of the expansion coefficients can be bypassed through the construction of an extended score vector. 
\cite{quli2006} extended QIF to varying coefficient models that are suitable for estimating models similar to equation \eqref{eq:nhanesMean}. 
Their approach, however, suffers from having to select knots, a difficulty that is exacerbated in high dimension where the number of choices grows. 
Moreover, their method is computationally burdened by the optimization of the QIF objective function over large amounts of data, rendering the method practically intractable for more than $200$ outcomes per individual and therefore unsuitable for intensively measured longitudinal outcomes of the type measured by wearable devices. 

To mitigate computational cost, a reasonable attempt at modeling time-varying effects estimates a functional regression model similar to equation \eqref{eq:nhanesMean} by ignoring dependence in the outcome. 
Estimation in functional regression consists of two dominant approaches: kernel and spline smoothers \cite[see][Chapters 3, 7-9]{ruppert2003}. 
Kernel-based methods suffer from the curse of dimensionality \citep{bellman1957, geenens2011}, bandwidth selection problems \citep{chen2015}, and ill-defined distance metrics \citep{beyer1999} in high dimensions.
Spline-based smoothers suffer from the variable selection problem in high dimensions and complexity of knot selection for certain basis functions \citep{ruppert2003}. 
Both kernel and spline smoothers have extensions for application with longitudinal outcomes \cite[see][Part III]{fitzmaurice2009} that account for dependence between observations for greater statistical efficiency. 
These extensions, however, can compound the previously mentioned complications \cite[see][Chapter 8.3 and references therein]{fitzmaurice2009}. 

In this paper, we develop the theory and methods for computationally and statistically efficient functional regression of intensively measured longitudinal outcomes. 
The primary technique is a division of the intensively measured longitudinal outcome into smaller blocks of dependent outcomes that can be analyzed in a computationally efficient manner. Regression models on small blocks of outcomes are estimated using spline basis approximations that fit naturally into this distributed framework.  
Our chief contributions are three-fold: 
(i) we propose methodology for distributed estimation of functional relationships between longitudinal outcomes and covariates that leverages within-subject correlation for optimal statistical efficiency; 
(ii) we develop the necessary theory for inference under a new constrained parameter space framework that encodes smoothness of the estimated functional parameter to any desired degree; and 
(iii) we extend this methodology for generalized linear models with both functional and fixed covariate effects.
Our key insight is to view knot locations as natural break points that partition the longitudinal outcomes into smaller longitudinal outcomes that are easily analyzed using simple polynomial basis function approximations. 
This leads to a natural formulation of the modeling objective as a data integration problem, in which subsets of data are analyzed and individual results are integrated following a specific rule of data heterogeneity and dependence.
The data integration formulation yields tremendous computational gains through its implementation in the distributed MapReduce paradigm.

Traditionally, distributed inference approaches split individuals into multiple blocks, estimate parameters within each block, and combine the parameter estimates using summary statistics from each block. 
Recent literature for aggregated estimating equations \citep{lin2011, tang2016} or combined confidence distributions \citep{xie2013, liu2015, zhou2017} assume observations across blocks are independent. 
Treating blocks of longitudinal outcomes as independent fails to account for the dependence between divided outcomes and leads to a substantial loss of statistical efficiency.

Relatively little work in combining estimators from dependent data blocks exists. Estimators proposed by \citet{chang2015} and \citet{li2020} are not robust to misspecification of the between-block correlation model.  
The (Doubly) Distributed and Integrated Method of Moments of \cite{hector2020a, hector2020b} provides the theoretical foundation for efficiently combining correlated estimates across dependent data sources. 
Their approach focuses on estimating homogeneous parameters of interest in the presence of heterogeneous nuisance parameters by combining estimating functions using the generalized method of moments \citep[GMM,][]{hansen1982}. 
Assuming homogeneous basis approximation parameters over each block is equivalent to the belief that the functional parameter is cyclical with period length equivalent to the block length. This assumption is unreasonable for wearable device data.
A more flexible approach would allow for heterogeneity of basis approximation parameters.

The estimated functional regression parameter is often assumed a smooth function over time. 
Specifically, separate estimates of $\bbeta(t)$ in each block of longitudinal outcomes exhibit undesirable discontinuities at partition break points, which is not consistent with our model or our intuition. 
To address this substantial difficulty, we encode smoothness of the parameter function directly in the estimation procedure through a constraint and penalty term added to a GMM objective function. By handling these constraints in the estimation rather than with computationally expensive post-processing procedures, we improve model interpretability and maximize computational efficiency. 
These additional constraints require us to develop asymptotic theory for constrained GMM estimators, which we achieve by projecting the parameter space into a lower dimension. 
We use this new theory to derive inferential properties of our proposed estimator of $\eta_1, \eta_2, \eta_3, \eta_4$ and $\beta_1(t)$. 
We then propose a smooth constrained meta (SCM) estimator asymptotically equivalent to the constrained GMM solution that delivers optimal statistical inference in exceptionally fast computational time. Finally, we establish full theoretical and computational scalability of our distributed approach in realistic data collection paradigms by establishing the asymptotic properties of our estimators when the number of longitudinal outcomes per participant diverges.

The remainder of this paper is organized as follows. We set up the problem, notation and a running example in Section \ref{s:notation}.
Section \ref{s:distStep} describes the proposed data partitioning, model construction and estimation, and efficient integration approach. 
In Section \ref{s:asymp}, we establish the asymptotic theoretical properties of the proposed distributed estimator. 
In Section \ref{s:impl}, we develop the statistically and computationally efficient smooth constrained meta (SCM) estimator. 
In Section \ref{s:sims}, we investigate the finite sample performance of our SCM estimator through simulations.
In Section \ref{s:NHANES}, we illustrate the application of our method with the analysis of physical activity counts in NHANES described by model \eqref{eq:nhanesMean}. 
Details, theorem conditions, proofs, and additional NHANES information can be provided upon request. 
Code for implementation, simulations, and data analysis is also available upon request. 

\section{Problem Setup} \label{s:notation}
Suppose we have $N$ independent observations $(\bY_i, \mX_i, \mZ_i)$, where $\bY_i=(Y_{i1},\dots,Y_{iM_i})^\top \in \R^{M_i}$ is the observed outcome for individual $i$ at $M_i$ time points, $\mX_i = (\bX_{i1} , \dots, \bX_{iM_i})^\top \in \R^{M_i \times q}$ a matrix of covariates with time-varying effects and $\mZ_i = (\bZ_{i1} , \dots, \bZ_{iM_i})^\top \in \R^{M_i \times p}$ a matrix of covariates with scalar effects.
We posit the mean model 
\begin{equation*}
    \mu_{im} = \E(Y_{im} \vert \bX_{im}, \bZ_{im}) = h\{ \bX^\top_{im}\bbeta(t) + \bZ^\top_{im}\boldeta \},
\end{equation*}
with $h$ a known canonical link function, $\boldeta = (\eta_{1},\dots,\eta_p)^\top$, $\bbeta(t)= \{\beta_1(t),\dots,\beta_q(t)\}^\top$ and,
$m=1, \ldots, M_i$. 
For now, we assume $M_i$ is finite for ease of exposition, and allow $M_i$ to diverge in Section \ref{ss:theory:diverging}.
We denote sampling times by $t_{i1},\ldots,t_{iM_i}$, and define $Y_i(t_{im}) = Y_{im}$ where we suppress the dependence on $t$, with $\bX_{im}$ and $\bZ_{im}$ defined similarly. 
Here, we have implicitly assumed that $\E(Y_{im} \vert \mX_{i}, \mZ_{i}) = \E(Y_{im} \vert \bX_{im}, \bZ_{im})$. 
By construction, $\mX_i$ is associated with outcome $\bY_i$ through functional parameters $\bbeta(t)$ and $\mZ_i$ is associated with $\bY_i$ through scalar, time homogeneous parameters $\boldeta$. 
Let $\mbox{Cov}( \bY_i \vert \mX_i, \mZ_i) \in \mathbb{R}^{M_i \times M_i}$, which does not depend on covariates. 
Denote by $\mI(k) \in \mathbb{R}^{k\times k}$ the identity matrix, $\mathit{0}(k_1,k_2) \in \mathbb{R}^{k_1 \times k_2}$ the matrix of zeros, 
and $\0_{k} \in \mathbb{R}^{k}$ the vector of zeros, 
for some positive integers $k,k_1,k_2$. 
We say that a function is of differentiability class $C^{v}$ if the first $v$ derivatives are continuous for some integer $v>0$. 
A function of differentiability class $C^0$ is continuous but has no continuous derivatives. 
We assume that $M_i$ is so large that estimation of $\btheta = \{\bbeta(\cdot), \boldeta\}$ using conventional approaches is computationally challenging or infeasible.

\textit{Broken Stick Example}:
The broken stick mean model is given by $\E (Y_{im})  = \beta_1(t) = \vert t \vert =t \{ \mathbbm{1}(t>0) - \mathbbm{1}(t \leq 0) \}$ for $t \in [-15,15]$ where $\mathbbm{1}(\cdot)=1$ if $\cdot$ is true. 
We will return to this example throughout to illustrate implementation details of our proposed approach.

\section{Distributed and Combined Estimation} \label{s:distStep}

\subsection{Knot Locations and Partitioning} \label{ss:knotsAndBlocks}

We propose using basis function expansions to estimate functional parameters $\bbeta(t)$; details are discussed in Section \ref{ss:localmodel}.
To reduce the computational burden of estimation with large $M_i$, we propose to view knot locations as natural breakpoints that partition the data into blocks over the index variable, $t$.
Let $\{ c_{j} \}_{j=0}^J$  denote the edges of a partition $\mathcal{P}=\{ \mathcal{P}_j\}_{j=1}^J$  
such that 
$\mathcal{P}_j = \{t: c_{j-1} \leq t < c_j \}$, where 
$\{t: t \in \mathcal{P}_j \}$ is non-empty for all 
$j$, $c_0 = \inf t$, $c_J = \sup t$, $j=1, \ldots, J$. 
The $j^{th}$ block of data is defined as
\begin{equation}
    B_j = \{ (Y_{im}, \bX_{im}, \bZ_{im}) : t \in \mathcal{P}_j  \} = \left[
    \begin{array}{ccc}
         \bY_{1j} & \mX_{1j} & \mZ_{1j} \\
         \vdots &  \vdots & \vdots \\
         \bY_{Nj} & \mX_{Nj} & \mZ_{Nj} \\
    \end{array} \right],
\end{equation}
with $\bY_{ij}= (Y_{im,j})_{m=1}^{M_{ij}} \in \mathbb{R}^{M_{ij}}$ and $\mX_{ij} \in \mathbb{R}^{M_{ij} \times q}$, $\mZ_{ij} \in \mathbb{R}^{M_{ij} \times p}$ respectively denoting the response vector and covariate matrices in block $j$ for individual $i$, where $\sum_{j=1}^J M_{ij} = M_i$. 
Correspondingly, let $\bmu_{ij} = \E(\bY_{ij} \vert \mX_{ij}, \mZ_{ij}) = (\mu_{im,j})_{m=1}^{M_{ij}}$. 
We distinguish between the terms ``partition'' and ``blocks'', which respectively refer to $\mathcal{P}$ and $\{ B_j \}_{j=1}^J$.

By partitioning data into blocks according to index variable $t$, we reduce the dimension of the largest analyzed unit of data.
This reduction enables us to use fewer basis functions to approximate $\bbeta(t)$. 
Data-driven approaches to selecting both the location and number of knots can increase the computational burden of functional regression with intensively measured outcomes \citep{johnson1990minimax, yao2008knot}. 
Indeed, in a distributed setting where the distributed estimation step depends on the knot selection, these data-driven procedures require multiple rounds of the distributed step, each with a candidate set of knots, which is computationally burdensome. 
Instead, we replace the selection of knot locations with a choice of partition edges $c_j$: we propose to choose partitions sufficiently small so that the functional parameters are well-approximated by a polynomial basis and that the computational burden within one partition set is small. 
In so doing, our approach does not require multiple iterations of the distributed step, delivering computational efficiency gains. We take this approach in the simulations of Section \ref{s:sims} and the data analysis of Section \ref{s:NHANES}. 
Domain specific knowledge may also be used to choose the location of partition edges. For example, these may correspond to the start and end of repetitions during a bout of exercise. 

\subsection{Local Model Specification} \label{ss:localmodel}

So far we have been vague in our description of the basis expansion for $\bbeta(t)$, suggesting only that it be simple.  
We propose a polynomial basis expansion for estimating each $\beta_u(t), t \in \mathcal{P}_j$, $u=1, \ldots, q$. 
Define the basis approximation for $\beta_u(t)$ in block $j$ as
\begin{equation} \label{e:basis-expansion}
    \beta_{ju}(t) = \sum_{d=0}^{d_u} \gamma_{j,ud}(t-c_{j-1})^d_{+} 
    = \bxi_{ju} (t)^\top \bgamma_{ju},
\end{equation}
where $(\cdot)_+ = \max(0, \cdot)$, $d_u$ is the degree of the polynomial basis expansion for parameter $\beta_u(t)$, $\bgamma_{ju} = (\gamma_{j, u0}, \dots, \gamma_{j, ud_u})^\top \in \mathbb{R}^{d_u}$ and $\bxi_{ju}(t)$ is the design vector of corresponding basis functions. 
Derivatives of the term $\bxi_{ju}(t)^{\top} \bgamma_{ju}$ with respect to 
$\bgamma_{ju}$ 
are functions of Vandermonde matrices \citep[][Chapter 7]{monahan2008}, which facilitates computations. 
When the values of $t$ are large, the scaled parameterization $(t-c_{j-1}) / (c_{j}-c_{j-1})$ can improve computational stability. 

The proposed basis expansion in equation \eqref{e:basis-expansion} appropriately models the functional form. 
In part, this is because we allow the model for $\beta_{ju}$ to vary in each block through the block-specific parameters $\bgamma_{ju}$. 
More importantly, this is a direct consequence of considering a small range of values of $t$ in each partition set, where the partitions are chosen small enough so that $\beta_{ju}(t)$ is simple.
Without the partitioning described in Section \ref{ss:knotsAndBlocks}, the expression for $\bbeta(t)$ may grow increasingly complex as more basis functions or knots are considered with increasing $M_i$. 
Though practically we recommend polynomial basis functions of degree $d_u=3$ for each block, our method generalizes and the theory holds for arbitrary basis functions.

In addition, a homogeneous parameter setting assumes $\bgamma_{ju} \equiv \bgamma_u$ and therefore $\beta_{ju}(t) \equiv \beta_u(t)$ for all $j$, which is equivalent to assuming $\beta_u(t)$ is cyclical with period length equal to the length of partition sets. 
Therefore, the homogeneous parameter setting is a special case of our approach. 

\textit{Broken Stick Example continued:} Recall $q=1$, $J=2$ with partition edges $c_0=-15$, $c_1=0$ and $c_2=15$ and $d_1=1$. Omitting the dependence of $\gamma_{j,ud}$ on $u$ since $q=1$, we parameterize $\beta_1(t)$ using
\begin{equation*}
    \beta_1(t) = 
    \begin{cases}
    \gamma_{1,0} + \gamma_{1,1} \{ t- (-15)\} & \text{ if } t \leq 0\\
    \gamma_{2,0} + \gamma_{2,1} t  & \text{ if } t > 0\\
    \end{cases}.
\end{equation*}
While other parameterizations are available, we see in Section \ref{ss:comb} that this parameterization simplifies calculations during the combination step.

\subsection{Distributed Estimation}

Efficient estimation of $\btheta_j=( \bgamma_{j1}^\top, \dots, \bgamma_{jq}^\top,\boldeta^\top )^\top$ depends on appropriately modeling and estimating the second-order moments of $\bY_{ij}$. 
Quadratic inference functions \citep{qulindsayli2000} are a powerful tool for bypassing the modeling and computational difficulties associated with estimating these nuisance parameters. 
QIF estimators are asymptotically efficient under mild regularity conditions when the working correlation structure is correctly specified.
Even under a misspecified working correlation structure, QIF estimators are still efficient within a general class of estimators \citep{qulindsayli2000, song2009} and more efficient than generalized estimating equations estimators \citep{liang1986}. 
Denote $\mR_{1,ij}, \dots, \mR_{W,ij} \in \mathbb{R}^{M_{ij}\times M_{ij}} $ a set of known basis matrices of zeros and ones. 
Then define the block $j$-specific extended score function,
\begin{equation*}
\begin{split}
\overline{\bg}_{j,N} (\btheta_j) 
    &= \frac{1}{N}\sum_{i=1}^{N} \bg_{ij}(\btheta_j; \bY_{ij}, \mX_{ij}, \mZ_{ij}) \\
    &= \frac{1}{N}
    \begin{pmatrix} 
    \sum_{i=1}^{N}\frac{\partial \bmu_{ij}(\btheta_j;\mX_{ij}, \mZ_{ij}) }{ \partial \btheta_j} ^\top \mA_{ij}^{-1/2} \mR_{1,ij} \mA_{ij}^{-1/2} \{ \bY_{ij} - \bmu_{ij}(\btheta_j;\mX_{ij}, \mZ_{ij}) \} \\
    \vdots \\
    \sum_{i=1}^{N} \frac{\partial \bmu_{ij}(\btheta_j;\mX_{ij}, \mZ_{ij}) }{ \partial \btheta_j} ^\top \mA_{ij}^{-1/2} \mR_{W_j,ij} \mA_{ij}^{-1/2} \{ \bY_{ij} - \bmu_{ij}(\btheta_j;\mX_{ij}, \mZ_{ij}) \} \\
    \end{pmatrix},    \\
\end{split}
\end{equation*}
where $\mA_{ij}$ is a diagonal marginal variance matrix and
$\mA_{ij}^{1/2} ( \sum_w a_{wj} \mR_{w,ij})^{-1} \mA_{ij}^{1/2}$ 
is the working covariance matrix of $\bY_{ij}$ conditional on $\mX_{ij}, \mZ_{ij}$. 
Many correlation structures such as independent, auto-regressive lag 1 or exchangeable are easily modeled with $W_j \leq 2$ and $\mR_{1,ij}$ the identity matrix. 
We refer the reader to \cite{qulindsayli2000} for details regarding more flexible correlation structure models. 
In practice, the choice of correlation structure is driven by {\em a priori} knowledge of the data type
\citep{fitzmaurice2004}. For example, longitudinal outcomes are frequently modeled using auto-regressive lag 1 correlation structure. 
As weighted sums over the observations $Y_{ij}$, the extended score functions automatically account for block-specific dimensions $M_{ij}$, thereby allowing block sizes $M_{ij}$ to differ without further adjustment.

The QIF estimator is defined as 
\begin{equation} \label{eq:QIF}
    \widehat{\btheta}_j = \arg \min \limits_{\btheta_j} \quad N\overline{\bg}_{j,N}(\btheta_j) ^\top \mC^{-1}_{j,N}(\btheta_j) \overline{\bg}_{j,N}(\btheta_j) = \arg \min \limits_{\btheta_j} \quad N Q_{j,N}(\btheta_j)
\end{equation}
where $\mC_{j,N}(\btheta_j) = N^{-1} \sum_{i=1}^N \bg_{ij} (\btheta_j; \bY_{ij}, \mX_{ij}, \mZ_{ij}) \bg_{ij}(\btheta_j; \bY_{ij}, \mX_{ij}, \mZ_{ij})^\top$ estimates the covariance of the extended score functions.
Remarkably, estimation of $\btheta_j$ using $\widehat{\btheta}_j$ in \eqref{eq:QIF} does not involve the nuisance parameters $a_{wj}$. This construction therefore bypasses estimation of the second-order moments, yielding computational efficiency gains. 
The minimization problem in equation \eqref{eq:QIF} is carried out via Newton-Raphson in a computationally efficient manner given the reduced data size in each block. 

The reader familiar with basis approximations will recall that they frequently involve smoothing to reduce over-fitting; this is discussed in Section \ref{ss:comb}. Finally, the QIF is a special case of GMM \citep{hansen1982}, a link we will exploit in the combination step.

\subsection{Combination Step} \label{ss:comb}

Let $\bgamma_u = (\bgamma_{1u}^\top, \dots, \bgamma_{Ju}^\top)^\top$, $u=1, \ldots, q$ and $\bgamma=(\bgamma_1^{\top},\dots ,\bgamma_q^{\top})^\top$. Define $\btheta = (\bgamma^\top, \boldeta^\top)^\top$ and denote its parameter space by $\bTheta$.
We propose a statistically efficient procedure for combining estimates $\widehat{\btheta}_j$ from each block to obtain an estimate of $\btheta$. 
Estimators across blocks are dependent due to the dependence between $\bY_{ij}$.
As in the distributed step, accounting for this dependence is essential to maximize statistical efficiency of the combined estimator. 
The important observation here is that the stacked extended score functions $\{ \overline{\bg}_{j,N}(\btheta_j)\}_{j=1}^J$ over-identify $\boldeta$, and just-identify $\bgamma$. 
The GMM \citep{hansen1982} is a state-of-the-art technique for estimating over-identified parameters.

Define the stacking operator $\mathrm{vec}(a_j) = (a_1,...,a_J)^\top$. 
We propose to stack the first-order moment conditions from each block by defining
\begin{align*}
\overline{\bG}_N( \btheta) &= \mathrm{vec} \bigl[ \nabla_{\btheta_j} \{ \overline{\bg}^\top_{j,N}(\btheta_j) \} C_{j,N}^{-1}(\btheta_j) \overline{\bg}_{j,N} (\btheta_j) \bigr].
\end{align*}
Again, $\overline{\bG}_N(\btheta)$ over-identifies $\btheta$ and we invoke the GMM to minimize a penalized quadratic form of $\overline{\bG}_N(\btheta)$:
\begin{equation} \label{eq:gmm1}
    \widehat{\btheta} = \underset{\btheta}{\mathrm{\arg \min}} \quad \overline{\bG}_N( \btheta)^\top 
    {{\mV}}^{-1}_N ( \widehat{\btheta}_{all}  )
    \overline{\bG}_N ( \btheta) + \lambda_N \btheta^\top \mD \btheta,
\end{equation}
where 
$\widehat{\btheta}_{all} = (\widehat{\btheta}_j)_{j=1}^J$, 
${{\mV}}_N ( \widehat{\btheta}_{all} )$ 
is the plug-in sample covariance of $\overline{\bG}_N( \btheta)$, $\lambda_N \geq 0$ a smoothing parameter and $\mD$ is a diagonal matrix with entries $1$ corresponding to $\bgamma$ and $0$ corresponding to $\boldeta$ similar to \cite{quli2006}. By stacking the first-order moment conditions rather than the extended score functions $\{ \overline{\bg}_{j,N}(\btheta_j)\}_{j=1}^J$ as in \cite{hector2021}, we have reduced the dimension of the GMM weight matrix ${\mV}(\widehat{\btheta}_{all})$ and therefore reduced the computational burden of its inversion. 
Nonetheless, for exceedingly large $J$ (e.g. $J \gtrsim 1000$), the plug-in sample covariance may remain computationally burdensome to invert.
In such cases, alternative weight matrices, such as a block tridiagonal matrix that accounts for the correlation between adjacent blocks, can be used without affecting the asymptotic properties in Section \ref{s:asymp}. 
This alternative weight matrix balances statistical and computational efficiencies by accounting for some temporal dependence with a weight matrix that is computationally fast to invert \citep{meurant1992review, ran2006inverses}.
Critically, the GMM formulation in \eqref{eq:gmm1} is amenable to the incorporation of the regularization term $\lambda_N \btheta^\top \mD \btheta$, which induces the desired smoothness of the estimated $\bbeta(t)$ by regularizing $\bgamma$.
Alternative $\mD$ and multiple $\lambda_N$ may be used for $q > 1$ to impose regularization separately for each $\beta_u(t)$, $u=1, \ldots, q$. The combination step is facilitated by the parametric basis expansion; in contrast, it is unclear how to select bandwidths or impose smoothing for  kernel-based estimators across dependent and disjoint blocks.

Despite the smoothness induced by the $L_2$ regularization term $\lambda_N \btheta^\top \mD \btheta$, the formulation in equation \eqref{eq:gmm1} does not guarantee continuity of the estimated $\bbeta(t)$ at partition edges. 
In fact, it results in undesirable discontinuities at time points $c_1, \ldots, c_{J-1}$.
We propose to encode continuity of the estimated $\bbeta(t)$ at partition edges directly into the optimization by constructing linear continuity constraints expressed in terms of the within-block parameters and basis functions. 
To enforce differentiability class $C^0$ of the functional parameter at partition edge $c_j$, constrain
\vspace{-0.5em}
\begin{equation*}
\begin{split}
    \beta_{j+1 \ u}(c_j) = \gamma_{j+1, u0}  
        & \doteq \beta_{ju}(c_j) = \sum_{d=0}^{d_u} \gamma_{j, ud}(c_j - c_{j-1})_+^d, \\
\end{split}
\end{equation*}
for 
$j=1,\dots,J-1$, $u=1, \ldots, q$. 
Across all blocks, this corresponds to
\begin{equation}
\label{e:cont-const}
    \beta_u(t) = \gamma_{1,u0} +\sum_{j=1}^J \sum_{d=1}^{d_u} \gamma_{j,ud} \min \bigl\{ \max(0, t-c_j), c_{j}-c_{j-1} \bigr\}^d,
\end{equation}
for $u=1, \ldots, q$. These constraints create an over-identification in our moment conditions for $\btheta$ in addition to the already over-identified moment conditions for $\boldeta$. Interestingly, the continuity constraint is a by-product of our distributed approach, which introduces additional parameters in each block $j$ to identify $\bbeta(t)$, $t \in \mathcal{P}_j$. The continuity constraint reconciles the necessary differences in parameterizations between the distribute and combine step. This is most easily seen by returning to the Broken Stick Example.

\textit{Broken Stick Example continued}: 
Continuity at the partition edge $c_1$ is imposed by constraining $\gamma_{1,0} + 15\gamma_{1,1} = \gamma_{2,0}$. 
In a non-distributed approach (NDA), $\beta_1(t)$ may be decomposed as $ \beta_1(t) = \gamma_{1,0} + \gamma_{1,1}\min(t,0) + \gamma_{2,1} \max(t, 0)$. 
In comparison, the distributed approach introduces an additional parameter $\gamma_{2,0}$ that, while necessary at the distributed step, becomes redundant at the combination step.

We introduce notation to formally incorporate these constraints.
Denote the constraint as $\mH_u \bgamma_u = \0$ for some matrix $\mH_u$. 
For $\beta_u(t)$ in differentiability class $C^0$, columns 
$(j-1)(d_u+1) + 1$ 
through column 
$j(d_u+1)+1$ of the $j^{th}$ row of $\mH_u$
are given by
$(1, c_{j} - c_{j-1}, \dots, ( c_{j} - c_{j-1})^{d_u}, -1)$, 
all else zero. 
Defining $\mH=\mathrm{bdiag}(\mH_1, \dots, \mH_q)$ where $\mathrm{bdiag}$ constructs a block diagonal matrix from its arguments, the set of continuity conditions for $\bbeta(t)$ is given by
$\mH \bgamma = \0$ . 
The dimension of $\mH$
depends on the number of continuous derivatives enforced. 
For example, for $\beta_u(t)$ in differentiability class $C^{v}$, we have $\mH \in \R^{(v+1) (J-1) \times \sum_{u=1}^q(d_u+1)}$. 
With polynomial basis functions, we note that $v < d_u$, since $v \geq d_u$ induces homogeneity of $\bgamma_{ju}$ across blocks. 

The combined smoothed GMM estimator that satisfies the continuity constraint is 
\begin{equation} \label{eq:combineSolve}
\begin{split}
        \widehat{\btheta}_{GMM} &= \arg \min \limits_{\btheta} \quad \overline{\bG}_N( \btheta)^\top {{\mV}}^{-1}_N (\widehat{\btheta}_{all}) \overline{\bG}_N( \btheta) + \lambda_N \btheta^\top \mD \btheta \\
    & \text{subject to:} \quad \mH \bgamma = \0. \\
\end{split}
\end{equation}
It is critical that the constraints encoded in $\mH$ be placed at the combination step: because parameter estimates are not shared across blocks, constraints cannot be enforced at the distributed step. 
To our knowledge, this constrained approach in GMM estimation has not been proposed before. Existing literature on linearly constrained GMM is in the context of hypothesis testing \citep{hall2005generalized}, a difference which has substantial consequences for the theoretical development in Section \ref{s:asymp}. 

As discussed in the Broken Stick Example, the distributed approach results in over-parameterization of the basis approximation for $\beta_u(t)$. 
The constraint at the combination step ensures the basis approximation parameters remain identifiable at the distributed step while reconciling redundant parameters at the combination step. We discuss implementation of the estimator in equation \eqref{eq:combineSolve} in Section \ref{s:impl}. 

\section{Asymptotic Theory} \label{s:asymp}
\subsection{Theory for Fixed $M_i$}
\label{ss:theory:fixed}

In this subsection, we establish inferential properties of the combined smoothed GMM estimator in equation \eqref{eq:combineSolve} with finite $M_i$. 
We extend the results to diverging $M_{i}$ in Section \ref{ss:theory:diverging}.
The objective function in equation \eqref{eq:combineSolve} is composed of two sets of moment conditions: the first, $\overline{\bG}_N$, results from our distributed approach; the second, imposed by $\mH \bgamma=\0$, reduces the dimension of the parameter space. Our primary theoretical contribution is to develop asymptotic theory for GMM estimators in the setting where a subset of the over-identifying moment conditions constrain parameters in the other moment conditions. 
Through careful construction of the reduced parameter space, we show that the constrained GMM estimator in equation \eqref{eq:combineSolve} can be re-expressed as an unconstrained GMM estimator, greatly facilitating the derivation of its inferential properties. 

Let $\btheta_* \in \R^{\mathrm{dim}(\btheta) - \mathrm{rank}(\mH)}$ be the set of parameters defined by systematically removing the parameters in $(\bgamma_{j1}^\top, \dots, \bgamma_{jq}^\top)^\top$ that are linear functions of $(\bgamma_{j'1}^\top, \dots, \bgamma_{j'q}^\top)^\top$ in previous blocks for $j' < j$. 
We refer to $\btheta_*$ as the constrained parameter of interest and denote its parameter space by $\bTheta_*$.
As an example, using a polynomial basis function approximation with $p \geq 1$, $q = 1$ (suppressing $u$ in the subscripts for all $\gamma_{j,ud}$), and a continuity constraint at partition edges (i.e. $C^0$ differentiability class), the constrained parameter of interest is $\btheta_*=\{ \gamma_{1,0}, ( \gamma_{j,d})_{d,j=1}^{d_1,J}, \boldeta\}$.
The constrained parameter of interest $\btheta_*$ has the smallest number of parameters required to identify the approximation for $\bbeta(t)$ and $\boldeta$ under the continuity constraint $\mH \bgamma = \0$. 

Formally, we define a matrix $\mRT$ that maps the constrained parameter $\btheta_*$ to $\btheta$ through $\btheta = \mRT \btheta_*$. The mapping $\mRT$ is formed using the continuity constraints in $\mH$. 
If no continuity constraints are imposed, then $\mRT$ is an identity matrix. 
To estimate $\bbeta(t)$ of differentiability class $C^0$, the mapping is $\mRT = \mathrm{bdiag}\{\mRT_1,\dots,\mRT_q, \mI(p) \}$ where, for $u=1, \ldots, q$, $\mRT_u = \prod_{j=1}^J \mRT_{uj}$ with $\mRT_{u1} = \mI \{ J(d_u+1) - (J-1) \}$ and
\begin{equation*}
\resizebox{\textwidth}{!}{%
$
    \mRT_{uj} = \begin{pmatrix}
    \mI\{ (d_u + 1)(j-2) \} & \mathit{0} \{ (d_u+1)(j-2),(d_u + 1) \}  & \mathit{0}\{(d_u+1)(j-2), d_u(J-j+1) \}  \\
    \mathit{0} \{ (d_u + 1), d_u(J-j+1) \} & \mI ( d_u + 1 ) & \mathit{0} \{ (d_u+1), d_u(J-j+1) \} \\
    \mathit{0} \{ 1,(d_u + 1)(j-2) \} & \{1, c_j - c_{j-1}, . . . ,(c_j - c_{j-1})^{d_u}\} & \mathit{0}\{1, d_u(J-j+1) \}  \\
     \mathit{0} \{d_u(J-j+1), (d_u + 1)(j-2) \} & \mathit{0} (d_u(J-j+1), d_u + 1 ) & \mI \{ d_u (J-j+1) \}  \\
\end{pmatrix},
$
}%
\end{equation*}
for $j > 1$. 
A map $\mRT$ for differentiability class $C^1$ can be provided upon request.

The construction of $\mRT$ enables us to define the unconstrained GMM estimator,
\begin{equation} \label{eq:combineSolve_smallDim}
\begin{split}
        \widehat{\btheta}_{*,GMM} &= \underset{\btheta_*}{\mathrm{\arg \min}} \quad \overline{\bG}_N(\mRT \btheta_*)^\top {{\mV}}^{-1}_N (\widehat{\btheta}_{all} ) \overline{\bG}_N(\mRT \btheta_*) + \lambda_N \btheta_*^\top \mRT^{\top} \mD \mRT \btheta_*.
\end{split}
\end{equation}
The unconstrained estimator in equation \eqref{eq:combineSolve_smallDim} is equivalent to the estimator in equation \eqref{eq:combineSolve} and therefore also satisfies $\mH \bgamma = \0$. The estimator in \eqref{eq:combineSolve_smallDim}, however, is easier to work with because we can develop theoretical properties over the reduced parameter space $\bTheta_*$ and extend the results to the full parameter space $\bTheta$. Casting the constrained estimator in equation \eqref{eq:combineSolve} as an unconstrained GMM estimator also allows it to benefit from the wealth of well-known properties of GMM estimators \citep{hansen1982}. 

Denote the true values of $\btheta$ and $\btheta_*$ as $\btheta_{0}$ and $\btheta_{*0}$ respectively.
Let $\bS_{\btheta_1}(\btheta_2) = \nabla_{\btheta_1} \overline{\bG}_N(\btheta_2)$, 
\begin{align*}
\bG_i(\mRT \btheta_*) &= \mathrm{vec} \bigl[ \E \{ \nabla_{\btheta_j} \overline{\bg}_{j,N}(\btheta_j)\}^{\top} \E \{ \mC_{j,N}(\btheta_j) \}^{-1} {\bg}_{ij}(\btheta_j; \bY_{ij}, \mX_{ij}, \mZ_{ij}) \bigr], \\ 
{\mV}_0(\mRT \btheta_*) &= \lim_{N \rightarrow \infty} \frac{1}{N} \sum_{i=1}^N \E\{\bG_i(\mRT \btheta_*) \bG_i(\mRT \btheta_*)^{\top} \} < \infty.
\end{align*}
We show in Theorems \ref{thm:gmm_cons} and \ref{thm:gmm_asynorm} that the estimator $\widehat{\btheta}_{GMM}$ in equation \eqref{eq:combineSolve} is consistent and asymptotically normally distributed.

\begin{theorem}[Consistency of $\widehat{\btheta}_{GMM}$]
\label{thm:gmm_cons}
Given conditions (A1) and (A2), and assuming $\lambda_N$ is $o(1)$, then $\widehat{\btheta}_{GMM} \overset{p}{\to} \btheta_0$ as $N \to \infty$.  
\end{theorem}

\begin{theorem}[Asymptotic normality of $\widehat{\btheta}_{GMM}$]
\label{thm:gmm_asynorm}
Given conditions (A1), (A2), and (A3), and assuming $\lambda_N$ is $o(N^{-1/2})$, then $\sqrt{N} ( \widehat{\btheta}_{GMM} - \btheta_0 ) \overset{d}{\to} \mathcal{N} (\0, \mRT \iSigma \mRT^{\top} )$ as $N \to \infty$ where
\begin{align*}
\iSigma = \bigl[
\E \bigl\{ \bS_{\btheta_*}(\mRT \btheta_{*0}) \bigr\}^{\top} {\mV}^{-1}_0(\mRT \btheta_{*0}) 
\E \bigl\{\bS_{\btheta_*}(\mRT \btheta_{*0}) \bigr\} 
\bigr]^{-1} .
\end{align*}
\end{theorem}

The proofs of Theorems \ref{thm:gmm_cons} and \ref{thm:gmm_asynorm} borrow from \cite{hall2005generalized} 
and more generally \cite{neweymcfadden1994}. 
For theoretical interest and uniformity with existing GMM theory, we allow the use of an arbitrary positive semi-definite matrix $\mW_N$ which converges in probability to a matrix $\mW$ in place of the efficient choice $\mV_N^{-1}(\widehat{\btheta}_{all})$. 
This general theory establishes consistency and asymptotic normality of estimators using weight matrices that are computationally fast to invert when $J$ is exceedingly large, such as the tridiagonal matrix discussed in Section \ref{ss:comb}.
We first show uniform convergence of the quadratic form given in \eqref{eq:combineSolve_smallDim} to 
$Q_0(\mRT \btheta_*) = \E \{\bG_i(\mRT \btheta_*) \}^{\top} \mW \E \{ \bG_i( \mRT \btheta_*) \} $
to establish consistency of $\widehat{\btheta}_{*,GMM}$. 
The proof then uses the continuous mapping theorem to derive properties of $\widehat{\btheta}_{GMM}$ from those of $\widehat{\btheta}_{*,GMM}$. 

The asymptotic normality in Theorem \ref{thm:gmm_asynorm} delivers inference for $\bbeta(t)$ through large sample confidence intervals using the plug-in estimator 
\begin{align*}
\widehat{\iSigma}  =  \bigl\{ 
\bS_{\btheta_*}^\top (\mRT \widehat{\btheta}_{*,GMM})
{\mV}^{-1}_N(\mRT \widehat{\btheta}_{*,GMM}) 
\bS_{\btheta_*} (\mRT \widehat{\btheta}_{*,GMM})
\bigr\}^{-1}
\end{align*}
of $\iSigma$. By construction, the estimator $\widehat{\btheta}_{GMM}$ is Hansen optimal, i.e. has variance at least as small as that of any other estimator defined using the same moment conditions. 
The estimator $\widehat{\btheta}_{GMM}$ remains consistent and asymptotically normal, though less efficient, when $\mV^{-1}_N (\widehat{\btheta}_{all})$ is replaced with an arbitrary positive semi-definite matrix $\mW_N$.

\subsection{Theory for Diverging $M_i$}
\label{ss:theory:diverging}

Wearable device data are collected at fixed temporal resolutions over long periods of time, possibly with no data collection termination endpoint. 
By the nature of this data collection paradigm, it is of particular theoretical interest to consider the setting where the number of observations $M_i, i=1, \ldots, N$ are not fixed but instead allowed to diverge.
To this end, we consider the infinite horizon setting with increasing domain asymptotics. 
In this setting, the dimension $M_{ij}$ in each block is fixed and additional blocks need to be created to accommodate the increasing dimension $M_i$ of $\bY_i$; in other words, the divergence of $J$ is driven by the fact that $M_i$ diverges. 
Consequently, the dimensions of $\bgamma$ and therefore $\btheta_*$ also grow as $M_i, i=1, \ldots, N$ tend towards infinity. 
We therefore introduce $\btheta_{*M}$, the parameter vector constructed similarly to $\btheta_{*}$, where the new notation signifies that its dimension increases when $M_i \to \infty, i=1, \ldots, N$ to accommodate the now diverging dimension of $\bgamma$. 

We show in Theorems \ref{thm:divJ_consist} and \ref{thm:divJ_asynorm} below that, under suitable conditions, the estimator $\widehat{\btheta}_{GMM}$ in equation \eqref{eq:combineSolve} is consistent and asymptotically normally distributed as $M_i \to \infty, i=1, \ldots, N$. Specifically, due to the growing dimensions of $\widehat{\btheta}_{GMM}$ and $\btheta_{*M}$, we establish asymptotic normality for a (finite) subset of the parameter estimators.
To this end, we define a selection matrix $\mathcal{H}_N \in \R^{h \times [p+J\{ \sum_{u=1}^q (d_u+1)  \}]}$, with $h$ fixed. 

\begin{theorem}[Consistency for Diverging J] \label{thm:divJ_consist}
    Given conditions (A1*)-(A4*), and assuming $\lambda_N$ is $o(1)$, $\left([p+J\{ \sum_{u=1}^q (d_u+1)  \}] \right) / N \to 0$, and $\left([p+J\{ \sum_{u=1}^q (d_u+1)  \}] \right)^6 / N \to 0 $, then 
    $\vert\vert \widehat{\btheta}_{M,GMM} - \btheta_{M0} \vert\vert = O_p(c_{1N})$ for $ c_{1N} = \big( [ p + J \{\sum_{u=1}^q (d_u + 1) \}  ] / N \big)^{1/2} $.
\end{theorem}

\begin{theorem}[Asymptotic Normality for Diverging J] \label{thm:divJ_asynorm}
    Given conditions (A1*)-(A6*), and assuming $\lambda_N$ is $o(N^{-1/2})$,  $q_N^2 /N \to 0$, 
    and $\left([p+J\{ \sum_{u=1}^q (d_u+1)  \}] \right)^7 / N \to 0 $,
    then 
    $N^{1/2} \mathcal{H}_N ( \widehat{\btheta}_{*M, GMM} - \btheta_{*M,0} ) \overset{d}{\to} N(\0, \iSigma_{\mathcal{H}})$ as $N,M_i \to \infty$, where 
    \begin{align*}
        \iSigma_{\mathcal{H}} = \mathcal{H}_N 
        \big[ 
        \E \{ \bS_{\btheta_{*N}}^\top (\mRT{\btheta}_{*N,0}) \}
            {\mV}_0(\mRT \btheta_{*0})^{-1}
        \E \{ \bS_{\btheta_{*N}} (\mRT{\btheta}_{*N,0}) \}
        \big]^{-1} 
         \mathcal{H}_N^\top.
    \end{align*}
\end{theorem}

The proofs of Theorems \ref{thm:divJ_consist} and \ref{thm:divJ_asynorm} borrow from \cite{fan2004nonconcave}, \cite{hector2020b}
and, more generally, \cite{van2000asymptotic}. 
As in Section \ref{ss:theory:fixed}, proofs are derived using an arbitrary positive semi-definite weight matrix $W_N$ in place of the efficient choice $\mV_N^{-1}(\widehat{\btheta}_{all})$.
Theorems \ref{thm:divJ_consist} and \ref{thm:divJ_asynorm} establish useful and surprisingly powerful results: our proposed distributed estimator permits efficient statistical inference for wearable device data collected in perpetuity.

\section{Implementation} \label{s:impl}

\subsection{Smooth Constrained Meta Estimator}

In practice, iteratively solving equation \eqref{eq:combineSolve_smallDim} requires computationally costly derivatives and inversions at each iteration of the minimization.
In addition, this optimization must be performed for each candidate value of $\lambda_{N}$, further exacerbating computation difficulties. 
We propose an asymptotically equivalent `one-step' estimator that directly estimates the constrained parameter $\btheta_{*}$ in Section \ref{s:asymp} at a substantially reduced computational cost. 
Specifically, we propose a `one-step' smooth constrained meta (SCM) estimator that combines the estimated $\boldeta$, and updates the block-specific estimates of $\bgamma$ while enforcing the continuity constraints.  The distributed implementation of the SCM estimator is implicit in the ``meta'' component of its name, as meta-estimators combine estimates from multiple data sources.

Let $\mDT = \mRT^\top \mD \mRT$. 
The SCM estimator is defined as
\begin{equation} \label{eq:one-step}
\begin{split}
    \widehat{\btheta}_{*, SCM} &= 
    \bigl\{ \mRT^\top \bS_{\btheta} 
     (\widehat{\btheta}_{all} ) ^\top 
    {\mV}^{-1}_N  (\widehat{\btheta}_{all} )
    \bS_{\btheta} (\widehat{\btheta}_{all} ) \mRT 
    + \lambda_N \mDT \bigr\}^{-1} \\
    & \sum_{j,j'=1}^J \bigl\{\mRT^\top \bS_{\btheta}
     (\widehat{\btheta}_{all} )^\top \bigr\}_{\cdot j} 
    \bigl\{{\mV}^{-1}_N  (\widehat{\btheta}_{all} ) \bigr\}_{j,j'}  
    \bigl\{\bS_{\btheta} (\widehat{\btheta}_{all} ) \bigr\}_{j',j'}
    \bigl( \widehat{\btheta}_{all} \bigr)_{j' \cdot} \quad, \\
\end{split}
\end{equation}
where $\{\mA\}_{\cdot j}$ and $\{\mA\}_{j \cdot}$ are the columns and rows respectively of a matrix $\mA$ corresponding to block $j$, and $\{\mA\}_{j,j'}$ are the rows and columns of a matrix $\mA$ corresponding to blocks $j$ and $j'$ respectively.
The SCM estimator $\widehat{\btheta}_{*, SCM}$ in equation \eqref{eq:one-step} is an estimator of the constrained parameter $\btheta_*$ and therefore satisfies $\mH \bgamma = \0$. A SCM estimator of $\btheta$ is given by 
$\widehat{\btheta}_{SCM}=\mRT \widehat{\btheta}_{*,SCM}$. Notably, $\widehat{\btheta}_{*,SCM}$ does not require any iterative optimization beyond the distributed step, so that it can be computed at very little cost.

Estimators $\widehat{\boldeta}_j$ are integrated similarly to the one-step estimator proposed by \cite{hector2020a}, whereas $\widehat{\bgamma}_j$ are efficiently updated using the information from integrating $\widehat{\boldeta}_j$ and the between-block correlations estimated by ${\mV}_N (\widehat{\btheta}_{all} )$. 
Thus, the combination step results in an efficiency gain across both homogeneous and heterogeneous parameters, similar to \cite{lin2011}. 
Additionally, the estimator proposed by \cite{hector2020a} is a special case of SCM when $\lambda_N=0$ and $q = 0$, implying $\mRT$ the identity. 

We show in Theorems \ref{t:one-step-consistency} and \ref{t:one-step-normality} that the estimator in equation \eqref{eq:one-step} possesses desirable statistical properties for inference.
\begin{theorem}[Consistency of $\widehat{\btheta}_{*, SCM}$]
\label{t:one-step-consistency}
Given conditions (A1) and (A2), and assuming $\lambda_N$ is $o(1)$, then $\widehat{\btheta}_{*, SCM} \overset{p}{\to} \btheta_{*0}$ as $N \to \infty$.  
\end{theorem} 

\begin{theorem}[Asymptotic normality of $\widehat{\btheta}_{*, SCM}$]
\label{t:one-step-normality}
Given conditions (A1), (A2), and (A3), and assuming $\lambda_N$ is $o(N^{-1/2})$, then $\sqrt{N} ( \widehat{\btheta}_{*, SCM} - \btheta_{*0} ) \overset{d}{\to} \mathcal{N} (\0, \iSigma )$ as $N \to \infty$
with $\iSigma$ given in Theorem \ref{thm:gmm_asynorm}.
\end{theorem}
We discuss point estimation of $\bbeta(t)$ and confidence intervals in Section \ref{ss:se_formula}.

Parallel results to Theorems \ref{thm:divJ_consist} and \ref{thm:divJ_asynorm} for $\widehat{\btheta}_{GMM}$ in Section \ref{ss:theory:diverging}, can be established for consistency and asymptotic normality of the SCM estimator $\widehat{\btheta}_{*,SCM}$ in equation \eqref{eq:one-step} as $M_i \to \infty$. 
These theorems are omitted here for space rather than lack of theoretical interest. 
These substantial theoretical developments require similar conditions on the growth rate of $J$. Remarkably, it follows from these results that the one-step estimator in equation \eqref{eq:one-step} and the GMM estimator in equation \eqref{eq:combineSolve_smallDim} are asymptotically equivalent, even as $M_i \to \infty$. 
Thus, the one-step estimator achieves Hansen optimal statistical efficiency at a fraction of the computational cost, even as the number of longitudinal outcomes diverges.

\subsection{Smoothing Parameter Selection} \label{ss:lambda}

Smoothing occurs at the combine step so that information from all blocks is used in updating the parameter estimates. 
Allowing regularization to occur during integration also smooths the entire function across all blocks rather than solely within each block.
Thus, smoothing and continuity constraints together ensure the desired continuity of $\bbeta(t)$. 
Let $\widehat{\btheta}_{*}$ denote a generic estimator that is a solution of either \eqref{eq:combineSolve_smallDim} or \eqref{eq:one-step}. To select $\lambda_N$, we extend the GCV statistic of \cite{ruppert2002} and propose to select $\widehat{\lambda}_N = \mathrm{argmin}_{\lambda_N} \mathrm{GCV}(\lambda_N)$ with
\begin{equation} \label{eq:gcv_criterion}
    \mathrm{GCV}(\lambda_N) = \frac{ \overline{\bG}_N^\top(\mRT \widehat{\btheta}_{*})
    {\mV}_N^{-1}(\mRT \widehat{\btheta}_{*})
    \overline{\bG}_N(\mRT \widehat{\btheta}_{*}) }
    { (1 - N^{-1} \mathrm{trace}
    [
    \{ 
    \ddot{Q}_N(\mRT \widehat{\btheta}_{*}) + \lambda_N \mD 
    \}^{-1}
    \ddot{Q}_N(\mRT \widehat{\btheta}_{*}) 
     ] )^2 }.
\end{equation}
where 
$\ddot{Q}_N(\mRT \widehat{\btheta}_{*}) = \bS_{\btheta}^{\top}(\mRT \widehat{\btheta}_{*})
    {\mV}_N^{-1} (\mRT \widehat{\btheta}_{*})
    \bS_{\btheta}(\mRT \widehat{\btheta}_{*})$.
In combination with the polynomial basis function expansion, the smoothing parameter mitigates the risk of overfitting.
Computing equations \eqref{eq:one-step} and \eqref{eq:gcv_criterion} for each candidate value of $\lambda_N$ 
lends itself naturally to parallelization. 

\subsection{Parallelization}

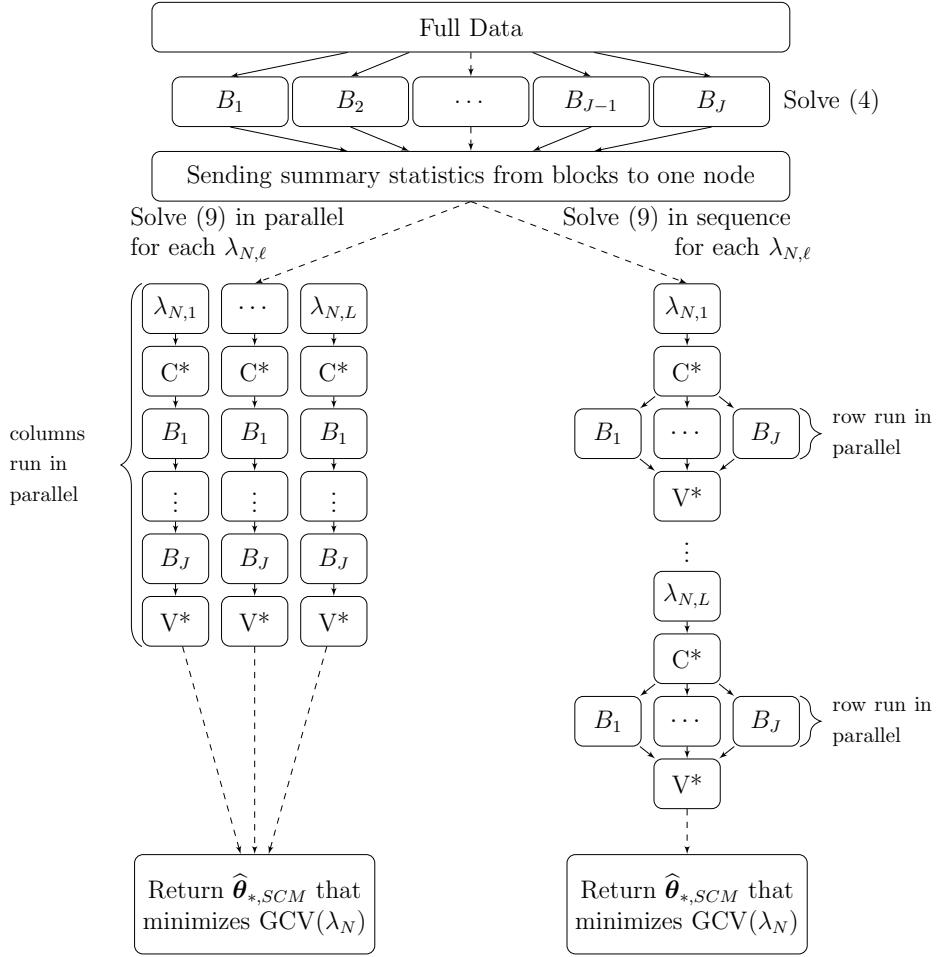
\begin{figure}[!ht]
    \tikzstyle{decision} = [diamond, draw, fill=none,  
    text width=4.5em, text badly centered, node distance=3cm, inner sep=0pt]
    \tikzstyle{data} = [rectangle, draw, fill=none, 
        text width=25em, text centered, rounded corners, minimum height=2em]
    \tikzstyle{block} = [rectangle, draw, fill=none, 
        text width=4em, 
        text centered, rounded corners, minimum height=2em]
    \tikzstyle{end_block} = [rectangle, draw, fill=none, 
    text width=9em,
        text centered, rounded corners, minimum height=4em]
    \tikzstyle{line} = [draw, -latex']
    \tikzstyle{cloud} = [draw, ellipse,fill=red!20, node distance=3cm,
        minimum height=2em]
    \tikzstyle{commentl} = [draw=none, fill=none, text width=10em, align=left]
    \tikzstyle{commentr} = [draw=none, fill=none, text width=10em, align=right]
    \tikzstyle{comment} = [draw=none, fill=none, text width=10em, text centered]
    \tikzstyle{empty_box} = [draw=none, fill=none, text width=6em, text centered]
    \tikzstyle{block2} = [rectangle, draw, fill=none, 
        text width=2em, text centered, rounded corners, minimum height=2em]

\begin{center}
\scalebox{0.8}{
    \begin{tikzpicture}[node distance = 2cm, auto]
        \node [data] (init) {Full Data};
    
        \node [block, below=4mm of init] (blockj) {$\dots$};
        \node [block, left of = blockj] (block2) {$B_2$};
        \node [block, left of = block2] (block1) {$B_1$};
        \node [block, right of = blockj] (blockJ1) {$B_{J-1}$};
        \node [block, right of = blockJ1] (blockJ) {$B_J$};
        \node [comment, right of = blockJ] (com1)  {Solve \eqref{eq:QIF}};
        
        \path[line,dashed] (init) -- (blockj.north);
        \path[line] (init) -- (block1.north);
        \path[line] (init) -- (block2.north);
        \path[line] (init) -- (blockJ1.north);
        \path[line] (init) -- (blockJ.north);
        
        \node[data, below=4mm of blockj] (int1) {Sending summary statistics from blocks to one node};
        
        \path[line,dashed] (blockj.south) -- (int1);
        \path[line] (block1.south) -- (int1);
        \path[line] (block2.south) -- (int1);
        \path[line] (blockJ1.south) -- (int1);
        \path[line] (blockJ.south) -- (int1);
        
        \node[empty_box, below=4mm of int1] (empty) {};
        \node[commentl, left=0mm of empty] (lambdapath) {Solve \eqref{eq:one-step} in parallel for each $\lambda_{N, \ell}$}; 
        \node[commentr, right=0mm of empty] (Jpath) {Solve \eqref{eq:one-step} in sequence \newline for each $\lambda_{N, \ell}$};

        \node[block2, below=2mm of lambdapath] (lamc) {$\cdots$};
        \node[block2, left=2mm of lamc] (lam1) {$\lambda_{N,1}$};
        \node[block2, right=2mm of lamc] (lamL) {$\lambda_{N,L}$};
        
        \node[rectangle, draw, fill=none, text width=2em, text centered, rounded corners, minimum height=2em, below=2mm of lamc] (combinelc) {C*};
        \node[rectangle, draw, fill=none, text width=2em, text centered, rounded corners, minimum height=2em, left=2mm of combinelc] (combinell) {C*};
        \node[rectangle, draw, fill=none, text width=2em, text centered, rounded corners, minimum height=2em, right=2mm of combinelc] (combinelr) {C*};
        
        \node[block2, below=2mm of combinell] (b11) {$B_1$};
        \node[block2, below=2mm of b11] (b1d) {$\vdots$};
        \node[block2, below=2mm of b1d] (b1J) {$B_J$};
        
        \node[block2, below=2mm of combinelc] (bc1) {$B_1$};
        \node[block2, below=2mm of bc1] (bcd) {$\vdots$};
        \node[block2, below=2mm of bcd] (bcJ) {$B_J$};
        
        \node[block2, below=2mm of combinelr] (bl1) {$B_1$};
        \node[block2, below=2mm of bl1] (bld) {$\vdots$};
        \node[block2, below=2mm of bld] (blJ) {$B_J$};
        
        \node[rectangle, draw, fill=none, text width=2em, text centered, rounded corners, minimum height=2em, below=2mm of bcJ] (combineVlc) {V*};
        \node[rectangle, draw, fill=none, text width=2em, text centered, rounded corners, minimum height=2em, left=2mm of combineVlc] (combineVll) {V*};
        \node[rectangle, draw, fill=none, text width=2em, text centered, rounded corners, minimum height=2em, right=2mm of combineVlc] (combineVlr) {V*};
        
        \node[rectangle, draw=none, fill=none, 
        text width=2em, text centered, rounded corners, minimum height=2em, below=2mm of combineVlc] (empty2) {};
        
        \node[rectangle, draw=none, fill=none, 
        text width=2em, text centered, rounded corners, minimum height=2em, below=2em of empty2] (empty3) {};

        \node[end_block, below of =empty3] (lambpar) {Return $\widehat{\btheta}_{*,SCM}$ that minimizes $\mathrm{GCV}(\lambda_N)$};
        
        \draw [decorate,decoration={brace,amplitude=10pt},xshift=-4cm,yshift=0pt]
        (combineVll.south west) -- (lam1.north west) node [black,midway,xshift=-.4cm,text width=4em] 
            {\footnotesize columns run in parallel};
        
        \draw[line, dashed] (int1.south) -- (lamc.north);
        
        \draw[line] (lamc) -- (combinelc);
        \draw[line] (combinelc) -- (bc1);
        \draw[line] (bc1) -- (bcd);
        \draw[line] (bcd) -- (bcJ);
        
        \draw[line] (lam1) -- (combinell);
        \draw[line] (combinell) -- (b11);
        \draw[line] (b11) -- (b1d);
        \draw[line] (b1d) -- (b1J);
        
        \draw[line] (lamL) -- (combinelr);
        \draw[line] (combinelr) -- (bl1);
        \draw[line] (bl1) -- (bld);
        \draw[line] (bld) -- (blJ);
        
        \draw[line] (bcJ) -- (combineVlc);
        \draw[line] (blJ) -- (combineVlr);
        \draw[line] (b1J) -- (combineVll);
        
        \draw[line, dashed] (combineVlc) -- (lambpar);
        \draw[line, dashed] (combineVlr) -- (lambpar);
        \draw[line, dashed] (combineVll) -- (lambpar);
        
        \node[block2, below= 2mm of Jpath] (lam1j) {$\lambda_{N,1}$};
        \node[rectangle, draw, fill=none, text width=2em, text centered, rounded corners, minimum height=2em, below=2mm of lam1j] (combinel1s) {C*};
        \node[block2, below=2mm of combinel1s] (lbd) {$\cdots$};
        \node[block2, left=2mm of lbd] (lb1) {$B_1$};
        \node[block2, right=2mm of lbd] (lbJ) {$B_J$};
    
        \draw [decorate,decoration={brace,amplitude=10pt},xshift=-4cm,yshift=0pt]
        (lbJ.north east) -- (lbJ.south east) node [black,midway,xshift=.4cm,text width=4.2em] 
            {\footnotesize row run in parallel};
        
        \node[rectangle, draw, fill=none, text width=2em, text centered, rounded corners, minimum height=2em, below=2mm of lbd] (combineVl1s) {V*};
        
        \node[rectangle, draw=none, fill=none, 
        text width=2em, text centered, rounded corners, minimum height=0em, below=0mm of combineVl1s] (vdots) {$\vdots$};
        
        \node[block2, below=0mm of vdots] (lamLj) {$\lambda_{N,L}$};
        
        \node[rectangle, draw, fill=none, text width=2em, text centered, rounded corners, minimum height=2em, below=2mm of lamLj] (combinel2s) {C*};
        
        \node[block2, below=2mm of combinel2s] (Lbd) {$\cdots$};
        \node[block2, left=2mm of Lbd] (Lb1) {$B_1$};
        \node[block2, right=2mm of Lbd] (LbJ) {$B_J$};
        \draw [decorate,decoration={brace,amplitude=10pt},xshift=-4cm,yshift=0pt]
        (LbJ.north east) -- (LbJ.south east) node [black,midway,xshift=.4cm,text width=4.2em] 
            {\footnotesize row run in parallel};
        
        \node[rectangle, draw, fill=none, text width=2em, text centered, rounded corners, minimum height=2em, below=2mm of Lbd] (combineVl2s) {V*};
        
        \node[end_block] at (combineVl2s|-lambpar) (Jpar) {Return $\widehat{\btheta}_{*,SCM}$ that minimizes $\mathrm{GCV}(\lambda_N)$};
        
        \draw[line, dashed] (int1.south) -- (lam1j.north);
        \draw[line] (lam1j) -- (combinel1s);

        \draw[line] (combinel1s) -- (lbd);
        \draw[line] (combinel1s) -- (lb1);
        \draw[line] (combinel1s) -- (lbJ);
        
        \draw[line] (lbd) -- (combineVl1s);
        \draw[line] (lb1) -- (combineVl1s);
        \draw[line] (lbJ) -- (combineVl1s);
        
        \draw[line] (lamLj) -- (combinel2s);
        \draw[line] (combinel2s) -- (Lbd);
        \draw[line] (combinel2s) -- (Lb1);
        \draw[line] (combinel2s) -- (LbJ);
        
        \draw[line] (Lbd) -- (combineVl2s);
        \draw[line] (Lb1) -- (combineVl2s);
        \draw[line] (LbJ) -- (combineVl2s);
        
        \draw[line, dashed] (combineVl2s) -- (Jpar);
        
    \end{tikzpicture}
} 
\end{center}        

    \caption{\label{fig:parallel_schema}
    Parallelization schemas: schema (i) is visualized on the left and schema (ii) is visualized on the right. $B_j$ indicates a QIF step on data from block $j$, C* indicates a combine step, and V* indicates estimation of a variance matrix and GCV criterion. 
    }
    
\end{figure}

We propose two parallelization schemas for implementation of the SCM estimator that leverage available computational resources for optimal computational efficiency. These schemas are visualized in Figure \ref{fig:parallel_schema}. 
Let $\lambda_{N,\ell}$ for $\ell=1,\dots,L$ be candidate values of $\lambda_N$.
The first distributed step computes $\widehat{\btheta}_j$ in equation \eqref{eq:QIF} in parallel for each block $j\in \{1, \ldots, J\}$. 
The summary statistics 
$\{ 
\widehat{\btheta}_j, 
\bg_{ij}(\widehat{\btheta}_j; \bY_{ij}, \mX_{ij}, \mZ_{ij}),  
\nabla_{\btheta_j} \overline{\bg}_{j,N}(\btheta_j)  \rvert_{\btheta_j=\widehat{\btheta}_j}
\} $ 
are then returned to a single computing node.
Then, one of two parallelizations occurs: 
\begin{enumerate}[label=(\roman*),topsep=0pt,itemsep=-1ex,partopsep=1ex,parsep=1ex]
    \item $L$ parallel processes are spawned for each $\lambda_{N,\ell}$, where each process solves \eqref{eq:one-step} and re-evaluates 
    $\overline{\bG}(\mRT \widehat{\btheta}_{*,SCM})$, $\bS_{\btheta}(\mRT \widehat{\btheta}_{*,SCM})$, and $\mV_N(\mRT \widehat{\btheta}_{*,SCM})$ 
    to compute
    \eqref{eq:gcv_criterion}, or
    \item for each $\lambda_{N,\ell}$ sequentially, one process computes $\widehat{\btheta}_{*,SCM}$ in \eqref{eq:one-step} and then computes re-evaluations of $\overline{\bG}(\mRT \widehat{\btheta}_{*,SCM})$, $\bS_{\btheta}(\mRT \widehat{\btheta}_{*,SCM})$, and $\mV_N(\mRT \widehat{\btheta}_{*,SCM})$ in parallel over $J$ nodes to compute \eqref{eq:gcv_criterion}.
\end{enumerate}
Computing the estimator for each $\lambda_{N,\ell}$ simultaneously following (i) requires $\max(J, L)$ computing nodes whereas 
sequential computation of the estimator for each $\lambda_{N,\ell}$ following (ii) uses at most $J$ computing nodes. 
We use both parallelization schemas in Section \ref{s:sims} to highlight the substantial computation gains of our approach over non-distributed approaches (NDAs) even when few computational resources are available. 
The most computationally advantageous schema will depend on whether $J$ iterations of QIF are faster than $L$ evaluations of \eqref{eq:one-step} and variance estimations.

\subsection{Large Sample Inference} \label{ss:se_formula}

For parsimony of notation, denote by $\widehat{\btheta}_*$ the SCM estimator $\widehat{\btheta}_{*,SCM}$ in equation \eqref{eq:one-step}. Theorem \ref{t:one-step-normality} suggests estimating the large sample covariance of $\widehat{\btheta}_*$ using
\begin{align}
\label{eq:SCM-sd}
    \widehat{\mathrm{Cov}} & (\widehat{\btheta}_*) = 
    \left\{ \Tilde{\mR}^\top \bS_{\btheta}(\mRT \widehat{\btheta}_*)^\top
    {\mV}^{-1}_N (\mRT \widehat{\btheta}_*)  \bS_{\btheta}(\mRT \widehat{\btheta}_*)\Tilde{\mR}  
    + \lambda_N \Tilde{\mD} \right\}^{-1}  \\
     & \Tilde{\mR}^\top \bS_{\btheta}(\mRT \widehat{\btheta}_*)^\top
     {\mV}_N(\mRT \widehat{\btheta}_*)^{-1}  
     \bS_{\btheta}(\mRT \widehat{\btheta}_*)\Tilde{\mR}  
     \left\{ \Tilde{\mR}^\top \bS_{\btheta}(\mRT \widehat{\btheta}_*)^\top
     {\mV}^{-1}_N(\mRT \widehat{\btheta}_*)  \bS_{\btheta}(\mRT \widehat{\btheta}_*)\Tilde{\mR}  + \lambda_N \Tilde{\mD} \right\}^{-1}. \nonumber
\end{align} 
Construction of pointwise large sample confidence intervals for the functional parameters $\beta_u(t)$ is achieved as follows.
Recall that for $t \in \mathcal{P}_j$, $\beta_{ju}(t) = \bxi_{ju}(t)^\top \bgamma_{ju}$. 
Define
\begin{align*}
\tilde{\boldsymbol{x}}_u &= \bigl\{ \0_{(u-1) J (d_u+1)}^\top,  
    \0_{(j-1)(d_u+1)}^\top,  
    \bxi_{ju}(t)^\top, 
    \0_{(J-j)(d_u+1)}^\top, 
    \0_{(q-u)J(d_u + 1)}^\top, 
    \0_{p}^\top \bigr\}^\top,
\end{align*}
where the vectors of zero align the basis functions with the corresponding basis parameters. 
Since $\beta_u(t)=\tilde{\boldsymbol{x}}^\top_u \mRT \btheta_*$, we estimate $\beta_u(t)$ with $\widehat{\beta}_u(t) = \tilde{\boldsymbol{x}}^\top_u \mRT
    \widehat{\btheta}_{*,SCM}$, 
and estimate $\boldeta$ with 
$\widehat{\boldeta} =
    \{\0_{J (d_u+1) q }^\top , 
    \mathbf{1}_{p}^\top \} 
    \mRT
    \widehat{\btheta}_{*,SCM}$,
with $\mathbf{1}_{p} \in \mathbb{R}^p$ a vector of ones. Suppressing the dimension of the zero and one vectors, we can construct large sample two-sided $100(1-\alpha)\%$ pointwise confidence intervals for $\beta_u(t)$ using 
\begin{equation} \label{eq:ci_method}
    \widehat{\beta}_u(t) \pm z_{1-\alpha/2} \sqrt{(\0^\top, \bxi_{ju}(t)^\top, \0^\top) \mRT ~\widehat{\mathrm{Cov}}(\widehat{\btheta}_{*,SCM}) \mRT^\top(\0, \bxi_{ju}(t), \0)},
\end{equation}
where $z_{1-\alpha/2}$ is the $100(1-\alpha/2)\%$ percentile of the standard normal distribution. 

The pointwise confidence interval construction can be extended to construct simultaneous confidence bands (SCB). 
With wearable device data, observations are densely measured and SCBs are constructed over a grid of (potentially unobserved) dense time points, $\mathcal{M}$, of dimension $M_{SCB}$.
In the literature, construction of these intervals typically proceeds using one of two approaches.
In the first approach, a wild bootstrap or simulation-based procedure re-samples perturbed observations and obtains model fits for each set of outcomes \citep{chang2017simultaneous}. 
In our setting, this method is computationally prohibitive due to the need to refit our model a large number of times. 
The second approach generates a number of pointwise confidence intervals and adjusts their width using a multiple comparison correction such as Bonferroni \citep{gu2014simultaneous, song2014simultaneous}, which can result in a loss of efficiency. 

In contrast, we show how to derive SCBs for $\bbeta_u(\mathcal{M})=\{\beta_u(t), t\in \mathcal{M}\} \in \R^{M_{SCB}}$ using Theorem \ref{t:one-step-normality}. 
The key insight is that, due to our parameterization, we can estimate $\bbeta_u(\mathcal{M})$ with $\widehat{\bbeta}_u(\mathcal{M})= \mathcal{B} \widehat{\btheta}_{*,SCM}$, where $\mathcal{B} = \mbox{bdiag}\{\tilde{\boldsymbol{x}}_u^\top(t_m) \mRT \}_{m \in \mathcal{M}}$, even at unobserved time points.
Manipulating the asymptotic distribution of $\widehat{\btheta}_{*,SCM}$, Theorem \ref{t:one-step-normality} shows that 
\begin{equation*}
    \sqrt{N} (\mathcal{B} \iSigma \mathcal{B}^\top)^{-1/2} \mathcal{B}(\widehat{\btheta}_{*,SCM} - \btheta_{*,0}) \overset{d}{\to} \mathcal{N} (\0, I). 
\end{equation*}
Define $\sigma^2_{B,mm} = (\mathcal{B} \iSigma \mathcal{B}^\top)_{mm}$ the $m$th diagonal element of $\mathcal{B} \iSigma \mathcal{B}^\top$. 
The asymptotic SCB for $\bbeta_u(\mathcal{M})$ at each point $t\in \mathcal{M}$ is then given by
$\widehat{\beta}_u(t) \pm \sigma_{B,mm} z_{1- \alpha/(2 M_{SCB})} / \sqrt{N}$.
Our SCBs exploit the covariance between two time points $t_m$ and $t_{m'}$ for efficient inference. Notably, our proposed critical values $z_{1-\alpha/(2 M_{SCB})}$ do not increase drastically with $M_{SCB}$: with $\alpha=0.05$ and 
$M_{SCB} \in \{10000, 1000000\}$, 
the SCB critical values are
$\{4.57, 5.45\}$, respectively.
Our critical values $z_{1-\alpha/(2 M_{SCB})}$ give slightly more precise inference for the same nominal $\alpha$ than  
those proposed by \citet{gu2014simultaneous}.
These critical values are also comparable to those in the functional data literature \citep{chang2017simultaneous, song2014simultaneous}. 

A further practical concern focuses on testing if the parameter $\beta_u(t)$ is a constant function of $t$. 
This can be expressed as a test of the hypotheses
$H_0: \gamma_{j,ud} = 0$ for all $(j,d) \ne (0,0)$ versus 
$H_A:$ at least one $\gamma_{j,ud} \ne 0$ for $(j,d) \ne (0,0)$. 
Equivalently, we can write 
$H_0: \mK^\top \btheta_* = \0$ and 
$H_A: \mK^\top \btheta_* \ne \0$ 
using a contrast matrix $\mK$.
Then, by Theorem \ref{t:one-step-normality}, we can construct a chi-squared test statistic for testing $H_0$ versus $H_A$ as $N \widehat{\btheta}_{*,SCM}^\top \mK (\mK^\top \iSigma \mK)^{-1} \mK^\top \widehat{\btheta}_{*,SCM}$. 
We reject $H_0$ when the observed test statistic is larger than the relevant quantile of the chi-squared distribution with degrees of freedom given by the rank of $\mK$.
Incidentally, this hypothesis testing framework can be deployed to test hypotheses for any set of linearly estimable functions of $\beta_u(t)$.

\section{Simulations} \label{s:sims}

\subsection{Objectives and Configuration}

We examine through simulations the empirical computational and statistical performance of the SCM estimator $\widehat{\btheta}_{*,SCM}$ in equation \eqref{eq:one-step}. 
Unless otherwise specified, we use cubic polynomial basis expansions to approximate the functional form of $\beta_u(t)$, $t\in \mathcal{P}_j$.
We show in various settings how the SCM estimator achieves nominal coverage of both scalar and functional parameters similar to NDA and clear computational superiority over NDA.
We demonstrate how analyses previously infeasible are rendered not only possible but computationally efficient through our approach.  
All simulations are conducted using \verb|R| on a Linux server. 
Run times are computed for both parallelization schemas in Figure \ref{fig:parallel_schema}. 
For ease of notation, we suppress $u$ in the subscripts for all $\gamma_{j,ud}$ when $q=1$.

\subsection{Broken Stick Model}

We consider the broken stick mean model 
$\E(Y_{im}) = \beta_1(t) = \vert t \vert$, $( t_{im} )_{m=1}^M = -15,-14, \dots, \\ 14, 15$, $i=1, \ldots, N$, 
$N=1000$.
Outcomes $\bY_i$ are generated from a multivariate normal distribution with exchangeable correlation structure (correlation $\rho = 0.7$ and variance $\sigma^2 = 10$). 
We partition data into two blocks using partition $\mathcal{P} = \{-15, 0, 15\}$ and estimate $\widehat{\beta}_1(t)$ in the differentiability class $C^0$, i.e. we impose continuity of $\widehat{\beta}_1(t)$ at $c_1$.
For computational stability, we use scaled $t$ for the SCM estimator as stated in Section \ref{ss:localmodel}. 
We compare our SCM estimator to the QIF estimator of \cite{qulindsayli2000}, a NDA.
Due to the scaled $t$, true parameter values for SCM and QIF are different. 
For SCM, we model
${\beta}_{j,1}(t) = \gamma_{j,0} + \gamma_{j,1} (t - c_{j-1})/(c_{j} - c_{j-1})$.
Thus, $\btheta_*=(\gamma_{1,0}, \gamma_{1,1}, \gamma_{2,1})$ and the true value is $\btheta_{*0} = (15,-15, 15)^{\top}$. 
The NDA parameterization is given by
\begin{align*}
    {\beta}_{1}(t) = \gamma_{1,0} + \gamma_{1,1} (t - c_0)/(c_2 - c_0) + \gamma_{2,1} \max \{ 0 , (t - c_1)/(c_2 - c_0) \},
\end{align*} 
and the true value is $\btheta_{0} = (15, -30, 60)^\top$. 
SCM estimation is parallelized over two computing nodes.
We do not consider regularization as both models are correctly specified and the dimension of the parameter is small. 
We report mean bias, mean asymptotic standard error (ASE) computed using equation \eqref{eq:SCM-sd}, empirical standard error (ESE) and 95\% confidence interval coverage probability (CP) for each parameter in $\btheta_{*} = (\gamma_{1,0}, \gamma_{1,1}, \gamma_{2,1})^\top$, averaged across 500 Monte Carlo samples in Table \ref{tab:sim1_sum_stat}. Because of the differences in values of $\btheta_{*0}$ and $\btheta_{0}$ that arise from scaling, we also report the relative asymptotic standard error (RASE = ASE/$\widehat{\gamma}_{j,d}$) in Table \ref{tab:sim1_sum_stat}.

\begin{table}  
\caption{\label{tab:sim1_sum_stat}Simulation results: Bias, ASE, ESE, RASE, and CP for SCM and QIF estimators in the broken stick model over 500 Monte Carlo samples.}  
\centering
\fbox{%
\begin{footnotesize}
    \begin{tabular}{rr rrrrr}
    Method & Parameter & Bias$\times 10^{-2}$ & ASE$\times 10^{-2}$ & ESE$\times 10^{-2}$ & RASE$\times 10^{-3}$ & CP \\
    \hline
    \multirow{3}{*}{SCM} & $\gamma_{1,0}$ & $-1.05$ & $8.71$ & $8.45$ & $5.81$ & $0.95$ \\
     & $\gamma_{1,1}$ & $-0.04$ & $3.68$ & $3.85$ & $-2.45$ & $0.94$ \\
     & $\gamma_{2,1}$ & $-0.02$ & $3.68$ & $3.83$ & $2.45$ & $0.94$ \\
    \multirow{3}{*}{QIF} & $\gamma_{1,0}$ & $-1.05$ & $8.71$ & $8.46$ & $5.81$ & $0.95$ \\
     & $\gamma_{1,1}$ & $-0.06$ & $7.36$ & $7.72$ & $-2.45$ & $0.93$ \\
     & $\gamma_{2,1}$ & $-0.02$ & $13.2$ & $14.9$ & $2.19$ & $0.94$ \\
    \end{tabular}
\end{footnotesize}
}
\end{table} 

The minimal bias in Table \ref{tab:sim1_sum_stat} indicates that SCM has good point estimation and is as accurate as QIF in this setting. 
The closeness between the asymptotic and empirical standard errors supports the use of the variance formula $\widehat{\mathrm{Cov}}(\widehat{\btheta}_{*,SCM})$ in the large sample setting. 
The RASE for SCM and QIF are almost identical, highlighting the fact that minimal statistical efficiency is lost by the distributed approach over a NDA. 
The CP for $\bgamma$ reaches nominal levels. Averaging across $t$, the CP for $\beta_1(t)$ is $95.6\%$ for both approaches.

\begin{figure}
    \centering
    \begin{subfigure}{0.49\textwidth}
        \centering
        \includegraphics[scale=1]{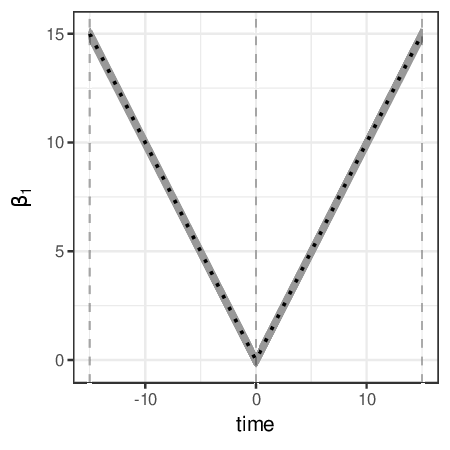}
        \caption{SCM}
        \label{fig:broken_stick_os}
    \end{subfigure}
    \hfill
    \begin{subfigure}{0.49\textwidth}
        \centering
        \includegraphics[scale=1]{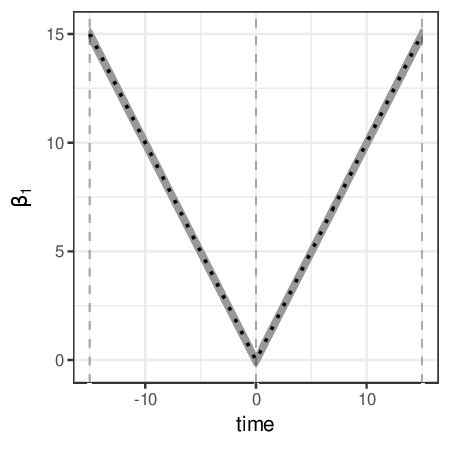}
        \caption{QIF}
        \label{fig:broken_stick_qif}
    \end{subfigure}
    \caption{\label{fig:broken_stick_graphs}
    Estimated $\widehat{\beta}_1(t)$ across 500 Monte Carlo samples with the true $\beta_1(t)$ overlaid in dotted black. Vertical dashed lines indicate the partition edges.
    }
\end{figure}

Computation for QIF takes an average of $9.31 (\mbox{sd}=0.438)$ seconds across the 500 Monte Carlo samples. 
SCM uses only one additional computing node and is considerably faster at $0.061 (0.058)$ seconds on average using schema (ii); schema (i) is not applicable as no values are considered for $\lambda_N$. 
Figure \ref{fig:broken_stick_graphs} shows point estimates $\widehat{\beta}_1(t)$ from all 500 Monte Carlo samples. In the Broken Stick Model, SCM delivers efficient statistical inference in substantially reduced computation time over a NDA.

\subsection{Known Distributed Parameterization}

We consider the mean model 
$\E(Y_{im}\vert X_{i}, Z_{i}) = X_{i} \beta_1(t) + Z_{i} \eta$
for 
$(t_{im})_{m=1}^M = 0,\dots, 99$,
$i=1,\dots,N$, $N=1000$, $\eta=6$ and 
\begin{equation*}
\begin{split}
    \beta_1(t) &= 1 + 2\frac{t}{20} - 3\frac{t}{20}^2 + 4\frac{t}{20}^3 + 
    5\left(\frac{t-20}{20}\right)_+^2 -2\left(\frac{t-20}{20}\right)_+^3
    -3\left(\frac{t-40}{20}\right)_+^2 \\
    &-10\left(\frac{t-40}{20}\right)_+^3  
    + 15\left(\frac{t-60}{20}\right)_+^2 + 20\left(\frac{t-60}{20}\right)_+^3 
    - 10\left(\frac{t-80}{20}\right)_+^2 + 5\left(\frac{t-80}{20}\right)_+^3. \\
\end{split}
\end{equation*}
We generate covariates following $X_{i} \sim \mathrm{Normal}(0,1)$ and $Z_{i} \sim \mathrm{Bernoulli}(0.5)$.
Outcomes $\bY_i$ are generated from a multivariate normal distribution with auto-regressive lag 1 correlation structure (correlation $\rho = 0.8$, variance $\sigma^2=100$ chosen large enough to provide visually distinguishable estimated functions).
By construction, $\beta_1(t)$ can be correctly specified by a cubic polynomial basis expansion at the combined step. 
We use partition 
$\mathcal{P}=\{0, 20, 40, 60, 80, 99\}$
for estimating $\beta_1(t)$ by SCM and constraints for the differentiability class $C^1$, i.e. we impose continuity of $\widehat{\beta}_1(t)$ and its first derivative at partition edges. 
We compare our approach to the penalized QIF (pQIF) of \cite{quli2006}. 
To make a fair comparison, we ensure both SCM and pQIF have approximately the same number of degrees of freedom by computing the pQIF estimator using knots placed every $10$ time points from $t=20$ to $90$. 
A grid of five values are considered for $\lambda_N$ for each method: $10^{-5}, 10^{-4},\dots,10^{-1}$ for SCM and $10^{-12}, \dots, 10^{-8}$ for pQIF. 

\begin{figure}[!ht]
    \centering
    \begin{subfigure}{0.45\textwidth}
        \centering
        \includegraphics[scale=0.9]{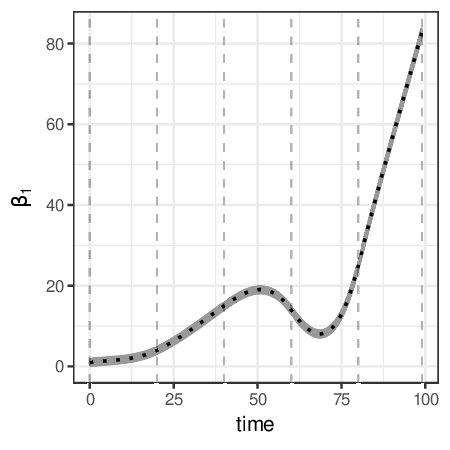}
        \caption{SCM}
        \label{fig:psim2_graphs_os}
    \end{subfigure}
    \hfill
    \begin{subfigure}{0.45\textwidth}
        \centering 
        \includegraphics[scale=0.9]{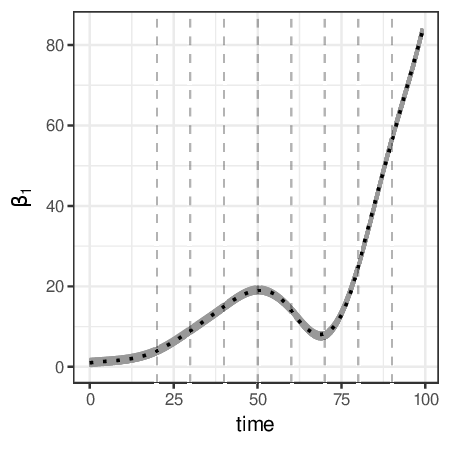}
        \caption{pQIF}
        \label{fig:psim2_graphs_trad}
    \end{subfigure}
    \caption{\label{fig:psim2_graphs}
    Estimated $\widehat{\beta}_1(t)$ across 500 Monte Carlo samples with the true $\beta_1(t)$ overlaid in dotted black. Vertical dashed lines indicate the partition edges in sub-figure (a) and knot locations in sub-figure (b).
    }
\end{figure}

Figure \ref{fig:psim2_graphs} shows the estimated $\widehat{\beta}_1(t)$ for each of $500$ Monte Carlo samples. 
The average of the empirical variance of the pQIF estimates in Figure~\ref{fig:psim2_graphs}~(b) is 0.0843, which is notably higher than the empirical variability of the SCM estimates in Figure~\ref{fig:psim2_graphs}~(a) (0.0788).  
Empirical $95\%$ confidence interval coverage (CP) of $\eta$ averaged over the 500 Monte Carlo samples is $96\%$ for SCM and $95\%$ for pQIF, whereas the average CP across all $\bgamma$ using SCM is $94\%$. 
There is no comparable measure for the $\bgamma$ using pQIF since the model is misspecified.
The average pointwise CP of ${\beta}_1(t)$ across the domain is $95\%$ for SCM and $86\%$ for pQIF.
The drop in coverage for pQIF may be explained in part due to model misspecification. 
It is possible pQIF could have achieved equivalent coverage with additional knot selection and penalization at the expense of greater computational burden and lower degrees of freedom. 
Both parallelization schemas for SCM require five computing nodes. 
Using schema (i) takes 
$1.57(\mbox{sd}=0.09)$ seconds,
and
using schema (ii) takes 
$1.72(0.09)$ seconds.
The pQIF average computation time of $304(156)$ seconds demonstrates how SCM offers a drastic reduction in the computation time even for relatively small $M_i=100$.
The SCM estimator is nearly 200 times faster than the NDA. We cannot overstate the computational scalability of our approach in this practical setting.
The undertaking of a knot selection procedure to achieve equivalent coverage would increase the computational burden of pQIF, further illustrating the computational superiority of SCM.

\subsection{Linear Model with Unknown Parameterization}

Additional simulations were performed in the linear model when $\beta_u(t)$ is misspecified by the basis approximation with both exchangeable and auto-regressive lag $1$ correlation structures. 
In these additional simulations, we show that the SCM estimator provides good point estimation, achieves nominal coverage of model parameters and is nearly $400$ times faster than a NDA. 
Our results demonstrate that, even for functional parameters without a closed form polynomial basis expansion, the SCM estimator provides can be efficiently used for statistical inference in large samples. 
Given the dramatic increase in computational burden of NDAs, we do not consider comparison to a NDA in the next simulation settings.

\subsection{Poisson Link Function}

\begin{figure}
    \centering
    \begin{subfigure}{0.45\textwidth}
        \centering
        \includegraphics[scale=0.9]{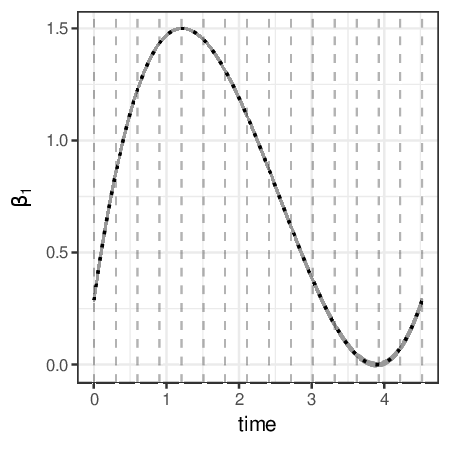}
        \caption{SCM Parameter Estimates}
        \label{fig:psimPois_os}
    \end{subfigure}
    \hfill
    \begin{subfigure}{0.45\textwidth}
        \centering
        \includegraphics[scale=0.9]{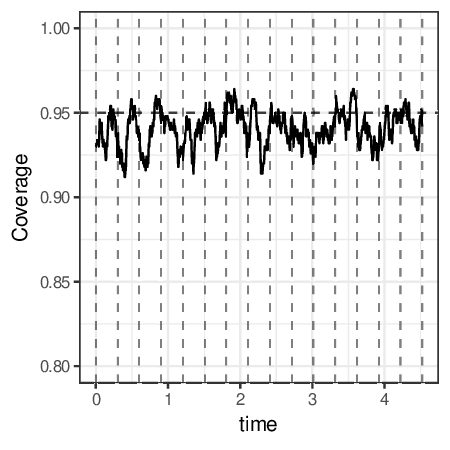}
        \caption{Coverage Probability}
        \label{fig:psimPois_cp}
    \end{subfigure}
    \caption{\label{fig:psimPois}
    Estimated $\widehat{\beta}_1(t)$ across $500$ Monte Carlo samples with the true $\beta_1(t)$ overlaid in dotted black. Vertical dashed lines indicate the partition edges. The CP is given in solid black with horizontal dashed reference line at the nominal level. 
    }
\end{figure}

We consider a Poisson regression setting that mimics one day of observed data from the NHANES application in Section \ref{s:NHANES}.
The mean model is $\log \{ \E(Y_{im} \vert X_{i}, Z_{i}) \} = X_{i} \beta_1(t) + Z_{i} \eta$ with $\beta_1(t)= 0.156 \{ t(t-\pi)(t-1438\pi/999 ) + 1.84 \}$ and $\eta=0.5$, 
$(t_{im})_{m=1}^M =0,0.003, \dots, 4.522$, $N=3000$, $M_i=1440$. Covariates are generated following $X_{i} \sim \mathrm{Uniform}(0.5, 5)$, and $Z_{i} \sim \mathrm{Bernoulli}(0.5)$.  
To generate Poisson outcomes, we first generate Gaussian errors with an autoregressive lag 1 correlation structure (correlation $\rho = 0.8$, variance $\sigma^2 = 1)$. 
We then apply the Poisson quantile function to the quantiles of the generated data, where the Poisson distribution has mean  $\exp \{ X_{i} \beta_1(t) + Z_{i} \}$. We partition data into $15$ blocks using partition $\mathcal{P}=\{0, 0.301, \dots, 4.221, 4.522\}$, and consider a grid of five values for $\lambda_N$: $10^{-5},10^{-4},\dots,10^{-1}$.  
Figure \ref{fig:psimPois} shows the estimated $\widehat{\beta}_1(t)$ across $500$ Monte Carlo samples and the empirical 95\% confidence interval coverage probability (CP) of $\widehat{\beta}_1(t)$. 
The estimates are nearly identical to the true $\beta_1(t)$ for each Monte Carlo sample, and CP for $\beta_1(t)$ averaged across $t$ and $\eta$ are $94\%$ and $95\%$, respectively. 
As anticipated in this difficult setting, the computation time increases: 
using schema (i) takes 
$358(\mbox{sd}=444)$ seconds on average, 
and
using schema (ii) takes 
$341(444)$ seconds. 
Run time is highly variable due to the non-linear link function.
These computation times are nonetheless incredibly small given the size of the data and complexity of the model. This simulation supports the use of SCM in the data application of Section \ref{s:NHANES}. 

\section{Application to the NHANES Data Set} \label{s:NHANES}

NHANES has collected phenotypic and health data on a representative sample of the U.S. population since the 1960s. Since 2003, NHANES has collected measures of physical activity using hip-worn accelerometers.
The accelerometer (ActiGraph AM-7164) detects, records and stores the intensity of movement (or counts) at a resolution of one minute for seven days. 
Participants were instructed to wear the monitors on their right hip for seven consecutive days, removing it only for sleep and to it keep dry. 
The analyzed data were accessed through the \verb|R| package \verb|rnhanesdata| \citep{leroux2019}. 

We consider a subset of $N=2772$ individuals from the 2003-04 and 2005-06 cohorts with $M_i=10080$ outcomes satisfying inclusion criteria available upon request. 
Our goal is to estimate the model in equation \eqref{eq:nhanesMean} and to carry out inference for the time-varying function $\beta_1(t)$, which quantifies the association between activity count and BMI, a surrogate of fitness.

The analysis dataset is large at $1.7$ GB of memory. Statistically efficient NDAs are rendered computationally infeasible by the need to estimate and invert the large $10080 \times 10080$ correlation matrix and model $\beta_1(t)$ over $10080$ time points. 
To overcome these difficulties, previous analyses of the NHANES data typically use summaries of activity count \citep[and references therein]{fuzeki2017}. 
Our estimation approach uses SCM for computationally and statistically efficient inference. 

We partition data into one hour blocks (e.g. 12:01 AM to 1:00 AM) resulting in $J=168$ blocks.
We estimate $\beta_1(t)$ in the differentiability class $C^1$, i.e. $\widehat{\beta}_1(t)$ has continuous first derivative at partition edges. 
The decaying dependence in $Y_{it}$ over time within each block is modeled using an autoregressive lag 1 working correlation matrix. 
The primary computational burden of SCM comes from the distributed step. 
We use a server with 64 cores available for the distributed step and consider eight values for $\lambda_N$: $10^{-1}, 10^{-0.5}, \dots, 10^{2.5}$.
Using schema (i) takes 
$49.38$ minutes with 
$88\%$ of the total computation time spent in the distributed step.  %
Using schema (ii) takes 
$54.17$ minutes with 
$80\%$ of the total computation time spent in the distributed step.

\begin{figure}[!ht]
    \centering
    \includegraphics[]{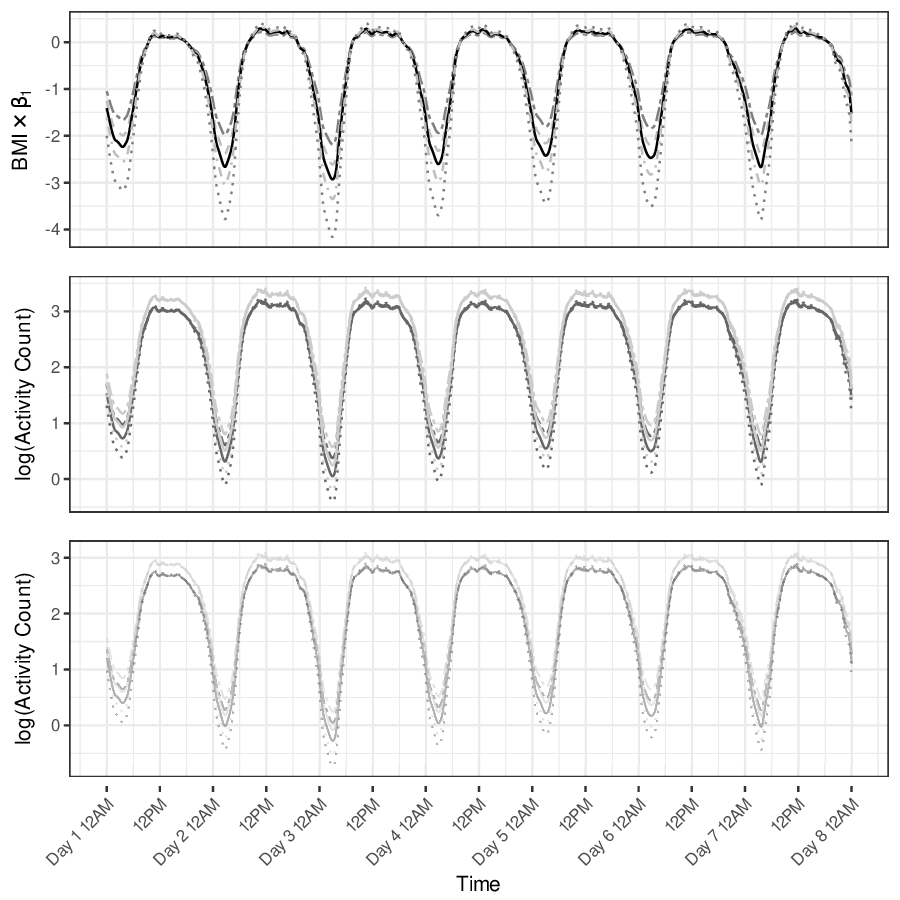}
    \caption{\label{fig:NHANES_1}
    [Top] Estimated form of $\beta_1(t)$ for various BMI quantiles. The double-dash, dashed, solid, dot-dash, and dotted lines represent the $5\%, 25\%, 50\%, 75\%, $ and $95\%$ BMI percentiles respectively. 
    [Middle] Estimated log activity count for individuals without mobility problems separated by sex (lighter indicates male). The double-dash, solid, and dotted lines represent the 
    $25\%, 50\%,$ and $75\%$ BMI and age percentiles respectively. 
    [Bottom] Estimated log activity count for individuals with mobility problems separated by sex (lighter indicates male). The double-dash, solid, and dotted lines represent the 
    $25\%, 50\%,$ and $75\%$ BMI and age percentiles respectively. 
    }
\end{figure}

Figure \ref{fig:NHANES_1} plots the estimated parameter, $\widehat{\beta}_1(t)$, and activity counts, 
\begin{align*}
    \log \widehat{\bY}_{quant}=\widehat{\eta}_{1} + \widehat{\eta}_{2} \mbox{Age}_{quant} + \widehat{\eta}_{3} \mbox{Sex} 
    + \widehat{\eta}_{4} \mbox{Mobility Problem} + \widehat{\beta}_1(t) \mbox{BMI}_{quant},
\end{align*} 
using $\widehat{\boldeta}$ the corresponding subset of $\widehat{\btheta}_{*,SCM}$ as described in Section \ref{ss:se_formula}, 
for different cross-sections of the study sample, where subscript $quant$ indicates a quantile of the corresponding variable, Sex $\in \{0,1\}$ (Sex~$=1$ for male sex) and Mobility Problem $\in \{0,1\}$ (Mobility Problem~$=1$ for presence of mobility problems). 
The estimated intercept is $\widehat{\eta}_{1} = -1.964$ with $95\%$ confidence interval (CI) $-1.984, -1.944$.
We estimate that the log activity count increases with each additional year of age by $\widehat{\eta}_{2} = 0.0955$ (95\% CI: $0.0945, 0.0964$).
Male sex is associated with an increase in the log activity count of $\widehat{\eta}_{3} = 0.195$ with 95\% CI $0.173, 0.217$. This increase in physical activity is visualized by the gap between greyscale lines in each panel of Figure~\ref{fig:NHANES_1}. 
Finally, mobility problems are associated with a decrease in the log activity count, with estimated mobility effect $\widehat{\eta}_{4} = -0.329$ (95\% CI $-0.363, -0.294$). This decrease is visually represented by comparison of the estimates in the middle and bottom panels of Figure~\ref{fig:NHANES_1}.
The estimated direction of sex associations is consistent with national health statistics in the United States \citep{HALLAL2012247, healthstat}. 
There is a clear pattern of decreasing activity around 9 P.M. daily and an increase in activity around 3 A.M. visualized in all panels of Figure \ref{fig:NHANES_1}.
The estimated $\widehat{\beta}_1(t)$ is negative from 7 P.M. to 7 A.M., indicating that larger BMI is associated with less movement during the non-working hours and slightly more movement during the working hours.

We have chosen to model the mean activity level of all included NHANES participants using equation \eqref{eq:nhanesMean}. An underlying assumption of this model is that, conditioned on covariates, all participants hold similar daily routines of physical activity. Subgroup analyses using SCM can be used when members of the study population have different daily routines, such as different sleeping or exercise times. Although not available in NHANES, daily diary data may be used to supplement models such as \eqref{eq:nhanesMean} with time-dependent covariates. In the past, granular analyses into daily physical activity patterns using intensive measurements have been computationally infeasible without assuming independence of the outcomes.
SCM is a promising method for analyses of similar data with the theoretical support for more complex analyses.

\section{Discussion} \label{s:discuss}

The SCM estimator delivers computationally and statistically efficient joint estimation of functional and scalar parameters under constraints and regularization in the presence of high dimensional correlation structures. We develop the theoretical framework for efficient statistical inference even as the number of outcomes diverges, and our simulations provide strong evidence of the feasibility of inference in large samples. 
The availability of two parallelization schemas provides a flexible framework for implementation under realistic computational resource constraints. 
The parallel framework provides massive computational gains over previous methods. 

We have suggested that the number of blocks $J$ be chosen to reduce the computational burden. 
In practice, placing partition edges near possible inflection points reduces the curvature to be captured by the within-block polynomial and may improve model fit. 
A fine partition with larger number of blocks $J$ decreases the computation burden, and can better estimate a complex curve by reducing the variability of the function estimated in each block, but the size of $J$ is limited by the sample size $N$.
Because asymptotic justifications for basis approximations rely on the remainder terms, i.e. gap between the approximation and true curve, going to $0$, finer partitions with cubic polynomials lead to increased accuracy.
The optimal choice of blocks ultimately depends on the true functional form of $\bbeta(t)$ and merits further investigation.  
Domain knowledge may also be used to choose the differentiability class of $\widehat{\bbeta}(t)$. 
Though not emphasized in this paper, partial homogeneity due to cyclical $\bbeta(t)$ or differing levels of continuity across partition edges are special cases of the proposed framework. 

The use of QIF in the distributed step avoids estimation of covariance parameters while maintaining score-like properties that can be leveraged to establish the combine step. The block estimators benefit from the well-known robustness and efficiency properties of the QIF. 
While the proposed methodology can be generalized beyond the QIF, 
alternative approaches using full likelihood or generalized estimating equations may incur additional computational costs and reduced robustness to covariance structure misspecification. 
At the combine step, a nonparametric optimal weight matrix ensures the QIF's desirable properties are transferred to and enjoyed by the final SCM estimator.
By combining the GMM and meta-estimation approaches, SCM achieves asymptotically optimal statistical efficiency but, crucially, avoids the significant computational burden of iterative optimization. 
The primary mechanism for this gain in computational efficiency is the translation of the linear constraint $\mH \bgamma = \0$ into moment restrictions and the subsequent construction of the unconstrained GMM estimator in equation \eqref{eq:one-step}. 
Our simulations show that the SCM estimator performs well under model misspecification. 

Our estimator automatically handles differing block sizes from varying dimensions $M_{ij}$. 
On the other hand, we have assumed that the number of individuals with observations in each block is a constant $N$. 
If 
the sample size in each block differs then some modification to the weights $1/N$ can be incorporated similar to the setting with independent blocks in \citet{hector2020b}. 
These modifications do not affect the consistency or asymptotic normality of the estimator. 
Relatedly, the SCM estimator is consistent when data are missing completely at random (MCAR): if, for example, data are missing due to MCAR drop-out, then the modification to the weights $1/N$ will yield an optimal estimator. 
When data that are not MCAR, further investigation into the theoretical properties of our estimator is required. 

The performance of the GMM is well known to deteriorate with small sample size $N$; see \cite{hansen1996} and others in the same issue. In these situations, sub-sampling following, for example, the approach of \cite{bai2012} may improve performance at a computational cost.

Possible applications of our SCM approach include analysis of data from personal activity trackers, continuous monitoring by clinical devices, and mHealth interventions. While we have emphasized the applicability to wearable device data, we anticipate our framework will be useful with spatial data applications.

\section*{Acknowledgements}
The authors are grateful for the insightful feedback from their colleague Dr. Marie Davidian. 

\section*{Data Availability}
The data that support the findings of this study are openly available in the rnhanesdata R package available at \verb|https://github.com/andrew-leroux/rnhanesdata|.

\section*{Funding}
This work was supported by a Faculty Research and Professional Development Award from North Carolina State University.

\appendix

\bibliographystyle{apalike}
\bibliography{bibliography}

\end{document}